\documentclass[a4paper,11pt]{article}
\pdfoutput=1 

\usepackage{jcappub} 

\usepackage[numbers,sort&compress]{natbib}

\newcommand{\be}{\begin{equation}}
\newcommand{\ee}{\end{equation}}
\newcommand{\bea}{\begin{eqnarray}}
\newcommand{\eea}{\end{eqnarray}}

\newcommand{\bra}[1]{\left\langle #1\right|}
\newcommand{\ket}[1]{\left| #1\right\rangle}

\DeclareMathAlphabet{\mathcal}{OMS}{cmsy}{m}{n}

\title{Slow-roll inflation in generalized scalar-torsion gravity}

\author[a]{Manuel Gonzalez-Espinoza,}
\author[a]{Giovanni Otalora,}
\author[a]{Nelson Videla}
\author[a]{and Joel Saavedra}

\affiliation[a]{Instituto de F\'{\i}sica, Pontificia Universidad Cat\'olica de Valpara\'{\i}so,\\Casilla 4950, Valpara\'{\i}so, Chile \label{addr1}}

\emailAdd{manuel.gonzalez@pucv.cl}
\emailAdd{giovanni.otalora@pucv.cl}
\emailAdd{nelson.videla@pucv.cl}
\emailAdd{joel.saavedra@pucv.cl}

\abstract{We study the dynamics of inflation in a generalized scalar-torsion gravity scenario by assuming a canonical scalar field non-minimally coupled to torsion with a Galileon-type self-interaction. 
After obtaining the field equations for a flat FRW background, we derive the second order action for both scalar and tensor perturbations to compute the power spectra of primordial fluctuations. 
As particular models, we studied at first, a power-law form of coupling function $F(x)=1+\xi x^{2}/2$, with $x\equiv \phi/M_{pl}$, and a monomial scalar field potential $V(x)=\lambda x^{n}/n$ which is ruled out by current observational data for $n\geq 2$. Under slow-roll approximation we obtain analytical expressions for the background as well as perturbative dynamics, and we show that the predictions of the model are consistent with current Planck 2018 constraints on the spectral index $n_{s}$ and the tensor-to-scalar ratio $r$ through the $n_s-r$ plane. Accordingly, this model is in agreement with current observational bounds only within the $95\%$ C.L. region in the case of chaotic quadratic inflation ($n=2$), whereas that for the other monomial potentials such as $n=4/3$, $n=1$ and $n=2/3$, it is found that they are even more favoured, overlapping their results with the $68\%$ C.L. region from last Planck data. Secondly, we studied a model in which the presence of both non-minimal coupling to gravity and the Galileon non-linear self-interaction $\gamma (\partial \phi)^2 \Box{\phi}$ leads to a suppression of the tensor-to-scalar ratio compared to those predicted in the standard scenario, then predicting
$0.024\lesssim r\lesssim  0.069$. This result allows us to reconcile chaotic quadratic inflation with current Planck data up to the $68\%$ C.L. region. In despite of this, next generation CMB experiments
such as BICEP3 or LiteBIRD are expected to put stronger constraints, making possible to either support this model or rule it out.}

\begin{document}
\maketitle
\flushbottom

\section{Introduction}\label{Introduction}
Although the Universe has undergone a decelerating expansion during the longest part of its lifetime, dominated first by radiation and then by matter, there are two accelerating phases in the history of the Universe.
The first accelerating phase corresponds to inflation \cite{Guth:1980zm,Starobinsky,Linde:1981mu}, which is widely accepted as the standard paradigm of the early Universe. The first reason is due to the fact that several long-standing puzzles of the Hot Big-Bang model, (HBB) such as the horizon, flatness, and monopole problems \cite{Liddle_Lyth2000}, find a natural explanation in the framework of inflationary Universe.  The simplest scenario explaining the physics of inflation is based on a canonical scalar field $\phi$, the inflaton, minimally coupled to gravity with a scalar potential $V(\phi)$ \cite{MukhanovBook}. In addition, and perhaps the most intriguing feature of inflation, is that it gives us a causal interpretation of the origin of the Cosmic Microwave Background (CMB) temperature anisotropies, while at the same time it provides us with a mechanism to explain the Large-Scale Structure (LSS) of the Universe, since quantum fluctuations during the inflationary era may give rise to the primordial density perturbations \cite{Baumann:2009ds,Liddle_Lyth2009,Mukhanov:1990me}.

From the viewpoint of quantum field theory in curved spacetime, a non-minimal coupling between the scalar field and curvature can naturally arise into the theory either by quantum corrections \cite{Linde:1982zj} or renormalizability requirements \cite{Freedman:1974gs,Freedman:1974ze,Birrell:1982ix}. Moreover, in the cosmological context,
a non-minimal coupling term accounts in modifying de dynamics of the inflaton field. Those
effects have been studied by several authors trough the literature. The non-minimal coupling term usually examined is $\xi \phi^2 R$, which was firstly considered for the new inflation scenario in Ref. \cite{Abbott:1981rg}, whereas that for chaotic inflation it was studied in Ref. \cite{Futamase:1987ua}, and later also in Refs. \cite{Tenkanen:2017jih,Linde:2011nh}. On the other hand, it has also been exhaustively investigated other different aspects of inflationary cosmology in the presence of non-minimal coupling such as for example the phase-space analysis \cite{Amendola:1990nn,Barroso:1991aj}, slow-roll approximation and conformal transformation techniques \cite{Faraoni:1996rf,Faraoni:2000wk,Faraoni:2000nt}, power-law inflation \cite{Tsujikawa:2000tm,delCampo:2015wma,Bairagi:2018ttm}, and  non-Gaussianities \cite{Qiu:2010dk,Nozari:2013mba,Piao:2002nh}. Also, the consequences of general non-minimal couplings for inflation have also been investigated in the framework of scalar-tensor theories, in the so-called extended and hyper-extended inflation \cite{La:1989za,La:1989pn,Holman:1990wq,Holman:1990jf,Wang:1991ww,Holman:1990tj,Laycock:1993bc,Steinhardt:1990zx,GarciaBellido:1990jz}. 

A very interesting scenario with a richer structure may also be obtained whether in addition to the non-minimal coupling term we also incorporate higher derivative quantum gravity corrections to the action such as for example a Galileon-type field self-interaction in the form $G(\phi,X) \Box \phi$, where $G$ is an arbitrary function of $\phi$ and $X\equiv -\partial_{\mu}\phi\partial^{\mu}{\phi}/2$ \cite{Deffayet:2010qz}. A self-interaction of this kind with $G\sim X$ arises naturally in the Dvali-Gabadadze-Porrati braneworld (DPG) model \cite{Dvali:2000hr} which corresponds to a non-linear interaction of the helicity-$0$ mode of the graviton \cite{Nicolis:2004qq}, being that more general functions $G(\phi,X)$ have also been considered for instance in Refs. \cite{Deffayet:2010qz,Silva:2009km,Kobayashi:2009wr,Kobayashi:2010wa,DeFelice:2010jn,DeFelice:2010gb}, and particularly in relation with $\alpha'$ corrections in low-energy effective string theory \cite{Metsaev:1987zx,Cartier:2001is}. These already mentioned large-distance modifications of General Relativity (GR) were firstly proposed to explain the late-time acceleration of the Universe without a cosmological constant. On the other hand, although single-field slow-roll inflation in GR provides us with the best fit to the data, considering alternative, non-standard scenarios, are motivated by the fact that certain scalar potentials for the inflaton coming from particle physics, such as the chaotic quadratic and quartic ones, are ruled out by current data. In this direction, in Ref. \cite{DeFelice:2011jm} the authors have investigated the chaotic inflation model in the context of general modified gravitational theories with non-minimal coupling term to curvature and Galileon self-interaction. In fact, the most general second-order scalar-tensor theory, the so-called Horndeski theory \cite{Horndeski:1974wa,Kobayashi:2019hrl}, which also includes non-minimal coupling models and generalized Galileon gravity, has been widely applied 
to account not only for the present accelerated expansion of the Universe but also in  studying the accelerated expansion of the early universe, such as in Refs. \cite{DeFelice:2011jm,DeFelice:2011zh}. Regarding the observational constraints on the Horndeski theory, by combing the gravitational wave event
GW170817, and the $\gamma$-ray burst GRB170817A, it has been possible to put strong constraints on the speed of Gravitational Waves (GWs), determining that
GWs propagate at the speed of light, with $-3\times 10^{-15}<c_{\textup{GW}}-1<7\times 10^{-16}$ \cite{Kobayashi:2019hrl}.
Accordingly, the viable subclass of the Horndeski
theory which satisfies $c_{\textup{GW}}=1$ is build only with k-essence, Galileon self-interaction and non-minimal coupling
terms, see for instance, Refs.\cite{Kobayashi:2019hrl} and \cite{Baker:2017hug,Sakstein:2017xjx}.

It is well known that gravity can be described in terms of torsion, in the context of so-called teleparallel equivalent of GR or simply Teleparallel Gravity (TG) \cite{Einstein,TranslationEinstein,Early-papers1,Early-papers2,Early-papers3,Early-papers4,Early-papers5,Early-papers6}. In this torsion gravity the dynamical variable is the tetrad field instead of the metric tensor $g_{\mu \nu}$, and the usual torsionless Levi-Civita connection of GR is replaced by the Weitzenb\"{o}ck connection, which has torsion but no curvature \cite{Aldrovandi-Pereira-book,JGPereira2,AndradeGuillenPereira-00}. The Lagrangian density of the theory is proportional to the scalar torsion $T$ which differs from the scalar curvature $R$ in a total derivative term and therefore the two theories are equivalent in the level of field equations \cite{Aldrovandi-Pereira-book,Arcos:2005ec}. Besides this equivalence, there are notable conceptual differences, being that the linear connection of TG is interchangeably identified with a purely spin connection and thus it arises as a classical gauge theory for gravitation based in the translation group, that due to existence of ``soldering" between the Minkowski tangent space (fiber) and the spacetime (base space), it becomes a non-standard gauge theory, keeping nevertheless a remarkable similarity to electromagnetism, also a gauge theory for an abelian group \cite{Pereira:2019woq,Aldrovandi-Pereira-book,Arcos:2005ec}. Moreover, following the same spirit of scalar-tensor theories, a natural extension for TG is a non-minimally coupled scalar-torsion theory where the scalar field matter source is non-minimally coupled not to the curvature scalar $R$ but to the scalar torsion $T$ \cite{Cai:2015emx,Bahamonde:2017ize}. An interesting aspect of this extension is that although TG coincides with GR at the level of field equations, this scalar-torsion theory is different from its counterpart based in curvature, that is to say, it belongs to a different class of gravitational modifications. Also, unlike what happens in scalar-tensor theories which are seen to be conformally related to Einstein's theory in absence of any matter field \cite{Clifton:2011jh}, here the non-minimal coupling to torsion cannot be removed by a conformal transformation and so this scalar-torsion theory does not have an equivalent minimally-coupled model \cite{Yang:2010ji,Sotiriou:2010mv,Wright:2016ayu}. A scalar-torsion theory with non-minimal coupling term in the form $\xi \phi^2 T$ has been originally applied to dark energy in Ref. \cite{Geng:2011aj,Geng:2011ka}, and then also, it was extended in Refs. \cite{Otalora:2013tba,Otalora:2013dsa}, for both, an arbitrary non-minimal coupling function $\phi^2\rightarrow f(\phi)$, and a tachyonic kinetic term for the scalar field, obtaining in both cases scaling attractors. More studies of the phase-space of dark energy were performed in Refs. \cite{Wei:2011yr,Xu:2012jf,Otalora:2014aoa,Sadjadi:2013nb,Fazlpour:2014qla,Skugoreva:2014ena}, while the dynamics of cosmological perturbations was addressed in Refs. \cite{Geng:2012vn,DAgostino:2018ngy}. Finally, applications of these scalar-torsion theories within the context of inflationary cosmology were made in Refs. \cite{Wu:2011kh,Wu:2016dkt}. 

The paper is organized as follows. In Section \ref{TG}, we give a brief introduction to TG. In section \ref{GS_torsion_gravity} we develop the framework of generalized scalar-torsion theory. In doing so, we calculate the modified field equations for a spatially Friedmann-Robertson-Walker (FRW) background. Then we derive the second order action for scalar and tensor perturbations to compute the power spectrum of primordial perturbations. In section \ref{Coupling} we study our first particular model, which only has non-minimal coupling term  of the form $F(x)=1+\xi x^{2}/2$, while the inflaton potential has a monomial form 
$V(x)=\lambda x^{n}/n$, with $x\equiv \phi/M_{pl}$. Here, by using the general formalism presented in section \ref{GS_torsion_gravity}, under the slow-roll approximation we solve analytically the background dynamics and derive the most important inflationary observables.
Then, we obtain the constraints on the parameters characterizing this model by comparing its predictions with Planck 2018 data. For our second model, in section \ref{Coupling_Galileon} we assume the presence of both, non-minimal coupling to torsion and Galileon self-interaction, along with the same expressions for the inflaton potential and coupling function used in \ref{GS_torsion_gravity}. We investigate the inflationary dynamics at background as well as perturbative level, contrasting the predictions of the model with current observational bounds, and so obtaining the physical constraints for the parameters. 
Finally, in Section \ref{Concluding_Remarks}, we summarize our findings and present our main conclusions and final remarks. 


\section{Teleparallel Gravity}\label{TG}

In teleparallel gravity (TG), a gauge theory for the translation group, the gravitational field is fully represented by the translational gauge potential which is identified as the non-trivial part of the tetrad field \cite{Aldrovandi-Pereira-book,JGPereira2,AndradeGuillenPereira-00,Arcos:2005ec}. Thus, one can use either the gauge potential or the tetrad field as the dynamical variable of the theory. The tetrad field $\mathbf{e}^{}_{A}(x^\mu)$ connects the spacetime metric $g_{\mu\nu}$ and the tangent space metric $\eta_{AB}$ thorough the local relation
\begin{equation}  \label{metrics}     
 g_{\mu\nu}=e^{A}_{~\mu}\,e^{B}_{~\nu}\,\eta_{AB}^{}\,,
 \end{equation} where $e^{A}_{~\mu}$ are the tetrad components in a coordinate base and then satisfying the orthogonality conditions  $e^{A}_{~\mu} e_{A}^{~\nu}=\delta^{\nu}_{\mu}$ and $e^{A}_{~\mu} e_{B}^{~\mu}=\delta^{A}_{B}$, being that $e_{B}^{~\mu}$ are the respective inverse components. The tangent space metric, $\eta_{AB}$ and $\eta^{AB}$, lowering and raising the Lorentz indices (Latin upper-case letters) is defined as the Minkowski metric $\eta _{AB}^{}=\text{diag}\,(-1,1,1,1)$, with $A= 0,\cdots,3$. On the other hand, the spacetime indices (Greek letters) vary from $0$ to $3$ and they are lowered and raised by the spacetime metric $g_{\mu\nu}$ and $g^{\mu\nu}$.

The action functional of TG is of the form
\be
S=\frac{M_{pl}^2}{2}\int{d^{4}x\,e\,{T}},
\label{action_TG}
\ee with $e=\det{\left(e^{A}_{~\mu}\right)}=\sqrt{-g}$, and $M_{pl}^2=\left(8\pi G\right)^{-1}$ is the reduced Planck mass. The torsion scalar
is defined as 
\bea
  T&=& S_{\rho}^{~\mu\nu}\,T^{\rho}_{~\mu\nu},\nonumber\\
&=& \frac{1}{4} T^{\rho \mu\nu}T_{\rho\mu\nu}+\frac{1}{2} T^{\rho\mu\nu} T_{\nu\mu\rho}-T_{\rho\mu}^{~~\rho} T^{\nu \mu}_{~~\nu},
\label{ScalarT}
 \eea
 where 
\begin{equation} \label{Def_Torsion}   
 T^{\rho}_{~\mu\nu}\equiv e_{A}^{~\rho}\left[\partial_{\mu}e^{A}_{~\nu}
 -\partial_{\nu}e^{A}_{~\mu}+\omega^{A}_{~B\mu}\,e^{B}_{~\nu}
 -\omega^{A}_{~B\nu}\,e^{B}_{~\mu}\right]\,,
\end{equation} are the components of torsion tensor, and
\begin{equation} \label{Superpotential}
 S_{\rho}^{~\mu\nu}=\frac{1}{2}\left(K^{\mu\nu}_{~~\rho}+\delta^{\mu}_{~\rho} \,T^{\theta\nu}_{~~\theta}-\delta^{\nu}_{~\rho}\,T^{\theta\mu}_{~~\theta}\right)\,,
\end{equation} is the so-called super-potential, with
\begin{equation}  \label{Contortion}
 K^{\mu\nu}_{~~\rho}= -\frac{1}{2}\left(T^{\mu\nu}_{~~\rho}
 -T^{\nu\mu}_{~~\rho}-T_{\rho}^{~\mu\nu}\right).
\end{equation} the contorsion tensor. The translational field strength of TG is the torsion tensor $T^{\rho}_{~\mu\nu}$, and as usually in gauge theories, the action is constructed by using quadratic terms in the field strength. Thus, the first term in the second equality of Eq. \eqref{ScalarT} is the usual one in gauge theories, whereas that the second and third terms are product of the soldering property of the spacetime manifold \cite{Pereira:2019woq,Aldrovandi-Pereira-book}. 

The spin connection of TG, $\omega^{A}_{~B \mu}$, is given by
\be
\omega^{A}_{~B \mu}=\Lambda^{A}_{~D}(x) \partial_{\mu}{\Lambda_{B}^{~D}(x)}.
\ee which represents only inertial effects of the frame and $\Lambda^{A}_{~D}(x)$ is a local (point-dependent) Lorentz transformation. It is just the connection that results after a local Lorentz transformation of the vanishing spin connection $\omega'^{A}_{~B \mu}=0$. For this connection one has that the curvature tensor vanishes identically
\bea
&& R^{A}_{~B\mu\nu}=\partial_{\mu}{\omega^{A}_{~B \nu}}-\partial_{\nu}{\omega^{A}_{~B \mu}}+
\omega^{A}_{~D \mu}\omega^{D}_{~B \nu} -\omega^{A}_{~D \nu}\omega^{D}_{~B \mu}=0,
\label{Curvature_Weitzenbock}
\eea whereas that the torsion tensor in Eq. \eqref{Def_Torsion} is non-vanishing. In this sense the spin connection $\omega^{A}_{~D \mu}$ can be seen as a kind of ``dual" of the spin connection of GR, which is a connection with vanishing torsion, but non-vanishing curvature. Furthermore, these two connections are the only two choices respecting the correct number of degrees of freedom
of gravitational field.

The linear connection corresponding to the spin connection $\omega^{A}_{~B \mu}$ is 
\be
\Gamma^{\rho}_{~\mu\nu}=e_{A}^{~\rho}\left(\partial_{\nu}{e^{A}_{~\mu}}+\omega^{A}_{~B \nu}e^{B}_{~\mu}\right),
\ee which is the so-called Weitzenb\"{o}ck connection. It is related to the Levi-Civita connection of GR, $\bar{\Gamma}^{\rho}_{~\mu\nu}$, through the equation
\be
\Gamma^{\rho}_{~\mu\nu}=\bar{\Gamma}^{\rho}_{~\mu\nu}+K^{\rho}_{~\mu\nu}.
\label{Gamma_relation}
\ee 

Substituting the relation \eqref{Gamma_relation} into Eq. \eqref{Curvature_Weitzenbock} and after taking the appropriate contractions it is easy to obtain 
\be
T=-\bar{R}-e^{-1} \partial_{\mu}{\left(e T^{\nu \mu}_{~~~\nu}~\right)},
\ee where $\bar{R}$ is the curvature of Levi-Civita connection. So, the action of TG, Eq. \eqref{action_TG}, differs from the Einstein-Hilbert action by a total divergence, and therefore the two theories are equivalent in the level of field equations.

By introducing a matter source, such a scalar field minimally coupled to gravity, we do not obtain any different result coming from the two theories. Nevertheless, in the non-minimal case the things are very different because there is a change in the way as the scalar field is coupled to gravity. In GR the scalar field is non-minimally coupled to the curvature scalar $R$, whereas that in TG the scalar field is non-minimally coupled to the torsion scalar $T$, and therefore, this change produces that the field equations do not coincide, which implies that the resulting theories are completely different. 

Below we are going to consider a generalized scalar-torsion theory and its implications for cosmic inflation.

\section{Generalized Scalar-torsion gravity}\label{GS_torsion_gravity}

\subsection{Background}

The relevant action is given by 
\begin{equation}
S=\int d^4x e \left[\frac{M_{pl}^2}{2} F(\phi) T+P(\phi, X)- G(\phi, X)\Box\phi \right],
\label{action1}
\end{equation}
where  $P$, $F$ and $G$ are arbitrary functions of $\phi$ and $X:=-(\partial\phi)^2/2$. This action extends the TG action \eqref{action_TG} to the case of a generalized scalar-torsion theory with a Galileon-type field self-interaction.

Now, we impose the standard homogeneous and isotropic background geometry, that is, we consider
\begin{equation}
\label{veirbFRW}
e^A_{~\mu}={\rm
diag}(1,a,a,a),
\end{equation}
which corresponds to a flat Friedmann-Robertson-Walker
(FRW) universe with metric 
\begin{equation}
ds^2=-dt^2+a^2\,\delta_{ij} dx^i dx^j \,,
\label{FRWMetric}
\end{equation}
where $a$ is the scale factor which is a function of the cosmic time $t$. Since the diagonal tetrad in Cartesian coordinates \eqref{veirbFRW} is a proper tetrad \cite{Krssak:2015oua}, we are going to take the vanishing spin connection, $\omega^{A}_{~B \mu}=0$, for our calculations. Thus, the background equations are given by
\bea
\label{00}
&& \xi_{00}\equiv P - 2 X P_{,X} - 6 H \dot{\phi} X G_{,X} + 2 X G_{,\phi}-3 H^2 M_{pl}^2 F=0,\\
&& \xi_{ii}\equiv P - 2 X G_{,\phi} - \dot{\phi} \dot{X} G_{,X} - 3 H^2 M_{pl}^2 F - 2 \dot{H} M_{pl}^2 F - 2 H M_{pl}^2 \dot{F}=0,
\label{ii} \\
&& \xi_{\phi}\equiv  P_{,\phi} - 2 P_{,\phi X} X - 18 H^2 X G_{,X} - 6 \dot{H} X G_{,X} + 2 G_{,\phi \phi} X + 3 H^2 M_{pl}^2 F_{,\phi}  \nonumber\\
&&- \Big[ 6 H X G_{,\phi X} + 3 H P_{,X}- 6 H G_{,\phi} \Big] \dot{\phi} -\Big[ P_{,X}+2 X P_{,XX}+6 H \dot{\phi} G_{,X}\nonumber\\
&& +6 H \dot{\phi} X G_{,XX}- 2 G_{,\phi} - 2 X G_{,\phi X} \Big] \ddot{\phi} = 0,
\label{phi}
\eea
where $H\equiv \dot{a}/a$ is the Hubble rate, where a dot represents derivative with respect to $t$. Also, a comma denotes derivative with respect to $\phi$ or $X$.

By eliminating the terms $P$ from Eqs. \eqref{00} and \eqref{ii}, one obtains the following relation
\bea
&& \epsilon=\delta_{PX}+\delta_{F}+3 \delta_{GX}+2\delta_{G\phi}+\delta_{\phi}\delta_{GX},
\label{epsilon1}
\eea where we have also introduced a set of slow-roll parameters
\bea
&&\epsilon\equiv -\frac{\dot{H}}{H^2},\:\: \delta_{\phi}\equiv \frac{\ddot{\phi}}{H\dot{\phi}},\:\: \delta_{F}\equiv\frac{\dot{F}}{H F},\:\: \delta_{PX}\equiv\frac{-X P_{,X}}{M_{pl}^2 H^2 F},\nonumber\\
&& \delta_{PXX}\equiv\frac{X^2 P_{,XX}}{M_{pl}^2 H^2 F},\: \delta_{G\phi}\equiv \frac{X G_{,\phi}}{M_{pl}^2 H^2 F},\: \delta_{G X}\equiv \dfrac{-\dot{\phi}X G_{,X}}{M_{pl}^2 H F}.
\label{SlowRoll-Parameters}
\eea
During inflation it is satisfied the condition $\epsilon\ll 1$, and hence, it is also required that all the slow-roll parameters defined above must be much smaller than the order of unity. Thus, at first-order approximation, we can put the expression \eqref{epsilon1} in the form 
\be
\epsilon=\delta_{PX}+\delta_{F}+3 \delta_{GX}+2\delta_{G\phi}+\mathcal{O}(\epsilon^2),
\label{epsilon2}
\ee where $\delta_{F},\delta_{PX},\delta_{GX},\delta_{G\phi}\ll 1$.
In what follows we compute the second-order action of scalar and tensor perturbations around the cosmological background in Eq. \eqref{FRWMetric}.

\subsection{Second-Order action}

In order to study primordial perturbations, our starting point is the ADM decomposition of the tetrad field \cite{Wu:2011kh,Wu:2016dkt}
\bea
&& e^{0}_{~\mu}=\left(N,\textbf{0}\right),\:\:\:\: e^{a}_{~\mu}=\left(N^{a},h^{a}_{~i}\right),\\
&& e_{0}^{~\mu}=\left(1/N,-N^{i}/N\right),\:\:\:\: e_{a}^{~\mu}=\left(0, h_{a}^{~i}\right),
\eea where $N^{i}=h_{a}^{~i}  N^{a}$, with $h^{a}_{~j} h_{a}^{~i}=\delta^{i}_{j}$, being $h^{a}_{~i}$ the induced tetrad field.

In the uniform field gauge, $\delta \phi=0$, a convenient ansatz for the fields is given by 
\be
N=1+\alpha,\:\:\:\: N^{a}=a^{-1} e^{-\mathcal{R}}\delta^{a}_{~i} \partial^{i}{\psi},\:\:\:\: h^{a}_{~i}=a e^{\mathcal{R}}\delta^{a}_{~j}\delta^{j}_{~i},
\label{Uniform_Field_Gauge}
\ee which gives the corresponding perturbed metric \cite{DeFelice:2011uc}
\bea
&& ds^2=-\left[\left(1+\alpha\right)^2-a^{-2}e^{-2\mathcal{R}}\left(\partial \psi\right)^2\right]dt^2+2\partial_{i}{\psi}dt dx^{i}+ a^2 e^{2\mathcal{R}}\delta_{i j} dx^{i} dx^{j}.
\eea
The additional degrees of freedom due to explicit violation of local Lorentz invariance in modified teleparallel gravity theories can be introduced by performing a Lorentz rotation of the frame $e^{A}_{~\mu}$ in Eq. \eqref{Uniform_Field_Gauge}, under the local Lorentz transformation $\Lambda^{A}_{~B}=\left(e^{\chi}\right)^{A}_{~B}=\delta^{A}_{~B}+\chi^{A}_{~B}+\frac{1}{2} \chi^{A}_{~C} \chi^{C}_{~B}+\mathcal{O}(\chi^2)$, while keeping fixed the spin connection of the background, $\omega^{A}_{~B \mu}=0$, in the form 
\be
e'^{A}_{~\mu}=\left(e^{\chi}\right)^{A}_{~B} e^{B}_{~\mu}=e^{A}_{~B}+\chi^{A}_{~B}e^{B}_{~\mu}+\frac{1}{2} \chi^{A}_{~C} \chi^{C}_{~B} e^{B}_{~\mu}+\mathcal{O}(\chi^2).
\ee
The Lorentz transformation matrix $\chi_{A B}=-\chi_{B A}$ can be then parametrized as 
\be
\chi^{0}_{~B}=\left(0,\chi_{b}\right),\:\:\:\: \chi^{a}_{~B}=\left(\chi^{a}, B^{a}_{~b}\right),
\ee where $\chi^{a}=\eta^{a b}\chi_{b}$ and  $B_{ab}=-B_{ba}$. Thus, one defines the spatial vector $\chi^{i}=h_{a}^{~i} \chi^{a}$ and the spatial antisymmetric tensor $B_{i j}=h^{a}_{~i} h^{b}_{~j} B_{a b}$. So, we also introduce the additional scalar mode $\beta$, the transverse vector mode $\chi^{(T)}_{i}$ and the (pseudo) vector mode $B_{i}$ in accordance with $\chi_{i}=\partial_{i}{\beta}+\chi^{(T)}_{i}$ and $B_{i j}=-B_{j i}=-\epsilon_{j i k} B^{k}$ \cite{Wu:2016dkt,Golovnev:2018wbh,Koivisto:2018loq,Li:2018ixg}.

From this point we follow closely the Maldacena's calculations \cite{Maldacena:2002vr}. The next step is to expand the action \eqref{action1} up to second order in perturbations which allows us to obtain 
\begin{eqnarray}
S^{(2)}&=& \int dt d^{3}x a^3\left[ \frac{1}{a^2} (2 w_1 \dot{\mathcal{R}} - w_2  \alpha ) \partial^2 \psi + 3 w_2 \alpha \dot{\mathcal{R}} \right. \nonumber
\\
&& \left. -  \frac{2 w_1 }{a^2} \alpha \partial^2 \mathcal{R} +\dfrac{1}{3} w_3 \alpha^2 - 3 w_1 \dot{\mathcal{R}}^2 + \dfrac{w_1}{a^2}(\partial \mathcal{R})^2 \right. \nonumber \\
&& \left. + 2 \left(  2 w_1 \dot{\mathcal{R}}  - w_2 \alpha \right) \partial^2 \beta +2 \dot{w}_{1} \mathcal{R} \partial^2 \beta  \right],  
\label{second_order}
\end{eqnarray}
where we have defined the functions
\begin{eqnarray}
w_1&=&  M_{pl}^2 F, \nonumber
\\
w_2&=& 4  \left[ \frac{M_{pl}^2}{2} H F + \dfrac{X^2}{\dot{\phi}} G_{,X} \right],\nonumber
\\
w_3&=& -3 X P_{,X} - 6 X^2 P_{,XX} + 6 X G_{,\phi} - 72 \dfrac{H}{\dot{\phi}} X^2 G_{,X}\nonumber
\\
&& +6 X^2 G_{,\phi X} - 36 \dfrac{H}{\dot{\phi}} X^3 G_{,XX} - 9 H^2 M_{pl}^2 F .
\end{eqnarray}
From the variation of action \eqref{second_order} with respect to $\alpha$ one obtains
\begin{equation}
\frac{1}{a^2} \partial^2 \psi =\frac{2 w_3}{3 w_{2}} \alpha +3\dot{\mathcal{R}}-\frac{2 w_1}{w_{2}} \frac{1}{a^2} \partial^2 \mathcal{R}-2 \partial^2 \beta,
\label{var1}
\end{equation}
whereas that the variation with respect to $\psi$ gives
\begin{equation}
\alpha = \dfrac{2 w_1}{w_2} \dot{\mathcal{R}},
\label{var2}
\end{equation}
Thus, from equations \eqref{var1} and \eqref{var2} we obtain the second order action for scalar perturbations
\begin{equation}
S_s^{(2)}= \int dt d^{3}x a^3 \ Q_s \left[\dot{\mathcal{R}}^2-\frac{c_s^2 }{a^2} (\partial \mathcal{R})^2 +2 \dfrac{\dot{w}_{1}}{Q_s} \mathcal{R} \partial^2 \beta \right], 
\label{SecOAction20}
\end{equation} where 
\be
Q_s=\dfrac{w_1}{3 w_2^2 } \left( 9 w_2^2 + 4 w_1 w_3 \right),
\label{Qs0}
\ee and the propagation speed of scalar mode $\mathcal{R}$ is
\be
c_s^2=\dfrac{3 \left(2 H w_1 w_2+ 4 \dot{w_1}w_2 -2 w_1 \dot{w_2}-w_2^2 \right)}{9 w_2^2 + 4 w_1 w_3}.
\label{Cs20}
\ee
In the second order action \eqref{SecOAction20} for scalar perturbations appears a new unconventional term with $\mathcal{R} \partial^2 \beta$ due to the explicit violation of local Lorentz symmetry. The scalar field $\beta$ does not necessary satisfies an Euler-Lagrange equation, but in order to maintain observer independence, such that $\delta{S}=0$ under observer local Lorentz transformations, it may be still necessary to impose the condition $\delta{S}/\delta \beta=0$, which may lead to a restricted geometry with undesirable results \cite{Kostelecky:2003fs,Bluhm:2014oua,Bluhm:2016dzm}. Something similar occurs in non-dynamical Chern-Simmons gravity, also a theory with explicit local Lorentz violation, where in order to evade the potential inconsistency between dynamics and geometry it is necessary to restrict the geometry by taking the Pontryagin density to vanishing, which also imposes severe restrictions on the dynamics of solutions \cite{Bluhm:2014oua,Bluhm:2016dzm,Alexander:2009tp}. In the case of non-minimally coupled scalar-torsion theories, if we impose the motion equation $\delta{S}/\delta \beta=0$ for the scalar $\beta$, then one must obtain the constraint $ F_{,\phi} \partial^2{\mathcal{R}}=0$, which leads us to $\partial^2{\mathcal{R}}=0$, in the case $F_{,\phi}\neq 0$. In the context of inflationary cosmology, it is a result that is not physically expected as it would imply that there are no nonzero-momentum solutions for the scalaron, and hence, no subhorizon scalar-perturbation mode would survive by the time of horizon crossing. This is currently an open issue in teleparallel scalar-torsion theories when applied to inflationary cosmology \cite{Wu:2016dkt}. One way in that one may try to solve this issue is restoring the local Lorentz invariance from the spacetime action \eqref{action1}, through a St\"{u}ckelberg-like mechanism \cite{Bluhm:2016dzm,Bluhm:2017pje,ArkaniHamed:2003uy}, or perhaps, by using Finsler geometries as suggested in Refs. \cite{Bonder:2018asb,Kostelecky:2011qz}. Nevertheless, an important point to be observed here is the following. In the FRW background the observer local Lorentz invariance is restored due to homogeneity and we can evolve the dynamics without any unforeseen. In fact, we have the problem that the field equations also depend on the spin connection, which, however, can be solved by introducing a reference frame and performing the scheme figured out in Ref.  \cite{Krssak:2015oua} for determining the appropriate spin connection. Hence, in the presence of non-minimal coupling, the field equations of background are exactly the same equations for a scalar-vector-tensor theory with spontaneous breaking of particle Lorentz symmetry due to a time-like vector field as in Ref. \cite{Kanno:2006ty} (see section \ref{Coupling}). Thus, the properties of these background equations change at the critical value $\phi_{c}$ defined by $f(\phi_{c})=1$, when parametrizing $F=1+f(\phi)$. For example, for $f(\phi)=\xi \phi^2/2$, it is easy to obtain $\phi_{c}=\sqrt{2/\xi}$, and thus, for $\xi\sim 10^{-3} M_{pl}^{-2}$ one has that $\phi_{c}\sim 45 M_{pl}$, which compared with the value at horizon crossing, $\phi_{*}\sim 15 M_{pl}$, it is seen that $\phi_{c}>\phi_{*}$, and therefore, the particle Lorentz violating stage with large contribution to inflation occurs in scales well deep inside the horizon. So, the modifications to the inflationary observables at the horizon crossing would be product of the reminiscent effect of this Lorentz violation stage \cite{Lim:2004js}. In this case, under slow-roll approximation, the Hubble parameter becomes constant, $H=\sqrt{\lambda/(3 \xi M_{pl}^2)}$, for the chaotic potential $V(\phi)=\lambda \phi^2/2$, even though the inflaton is rolling down the potential, as consequence of (particle) Lorentz violation \cite{Kanno:2006ty}. Moreover, observer independence is restored at the linear perturbations level, when taking the limit of very strong non-minimal coupling, $f(\phi)\gg 1$, as the background coefficient that accompanies the term $\mathcal{R} \partial^2 \beta$ in action \eqref{SecOAction20}, decays rapidly as $a^3 \dot{\phi} \phi \sim a^{3-4 \xi M_{pl}^2}$ for $a=(-\tau H)^{-1} \rightarrow 0$ \cite{Kanno:2006ty}. It is also straightforward to see that the limit of strong Galileon coupling does not change this result, as the Hubble rate continues to be a constant that depends only on $\xi$ and $\lambda$. On the other hand, at superhorizon scales the constrained geometry, with $\partial^2{\mathcal{R}}=0$, it is consistent with the inflationary picture in the limit of wavenumber $k\rightarrow 0$. In this large-scale limit the curvature perturbation $\mathcal{R}$ freezes after horizon crossing, and thus, by performing the standard quantization procedure from the action 
\begin{equation}
S_s^{(2)}= \int dt d^{3}x a^3 \ Q_s \left[\dot{\mathcal{R}}^2-\frac{c_s^2 }{a^2} (\partial \mathcal{R})^2 \right], 
\label{SecOAction2_Standard}
\end{equation} the power spectrum can be calculated as the two-point correlation function $\bra{0}\mathcal{R}(\tau, \vec{k}_{1}) \mathcal{R}(\tau, \vec{k}_{2})\ket{0}$, at some conformal time $\tau$ after the horizon exit, or, at the end of inflation $\tau\approx 0$ \cite{DeFelice:2011uc}. So, in order to proceed forward, we are going to neglect the observer Lorentz violating term in action \eqref{SecOAction20}, and let us take the action \eqref{SecOAction2_Standard} for evolve the dynamics of curvature perturbation in the usual way.

By expanding in terms of the slow-roll parameters \eqref{SlowRoll-Parameters}, the quantities $Q_{s}$ and $c_{s}^2$, in Eqs. \eqref{Qs0} and \eqref{Cs20}, respectively, are rewritten as follows
\bea
\label{Qs}
&& Q_{s}\simeq M_{pl}^2 F\Big[\delta_{PX}+6 \delta_{GX}-6 \delta_{GXX}+2 \delta_{G \phi}-2 \delta_{PXX}\Big],\\
&& c_{s}^2\simeq \frac{\delta_{PX}+2 \delta_{F}+4 \delta_{GX}+2 \delta_{G\phi}}{\delta_{PX}+6 \delta_{GX}-6 \delta_{GXX}+2 \delta_{G\phi}-2 \delta_{PXX}}.
\label{csSlow}
\eea Here we may also introduce the following parameter:
\be
\epsilon_{s}\equiv\frac{Q_s c_{s}^2}{M_{pl}^2 F}=\frac{w_{1} \left(2 H w_1 w_2+ 4 \dot{w_1}w_2 - 2 w_1 \dot{w_2}-w_2^2 \right)}{M_{pl}^2 F w_{2}^2}.
\ee Then, in terms of the slow-roll parameters we find
\bea
\epsilon_{s}&=& \epsilon+\delta_{F}+\delta_{GX}+\mathcal{O}(\epsilon^2),\nonumber\\
&=& 2 \delta_{F}+4\delta_{GX}+2 \delta_{G\phi}+ \delta_{PX}+\mathcal{O}(\epsilon^2).
\label{epsilons2}
\eea where in the first equality the term $\delta_{G\phi X}=X^2 G_{\phi X}/(M_{pl}^2 F H^2)$ has also been neglected because it is higher than the first order \cite{DeFelice:2011uc}. Also, in the second part of this equation we have used Eq. \eqref{epsilon2}. Finally, the presence of ghosts and Laplacian instabilities may be avoided if one is restricted only to cases $Q_s>0$ and $c_{s}^2>0$ \cite{DeFelice:2011uc,DeFelice:2011zh}. 

By varying the second order action \eqref{SecOAction2_Standard} in terms of $\mathcal{R}$ we obtain the equation of motion for the curvature perturbation $\mathcal{R}$ . Through the standard canonical quantization procedure applied to curved space-times \cite{Bardeen:1980kt,Kodama:1985bj,Mukhanov:1990me,Bassett:2005xm}, we obtain the power spectrum for curvature perturbation in the form
\bea
&&\mathcal{P}_{s} \equiv \frac{H^2}{8 \pi^2 Q_s c_{s}^3} = \frac{H^2}{8 \pi^2 M_{pl}^2 F c_{s} \epsilon_s }, \nonumber\\
&& \simeq \frac{H^2\left(\delta_{PX}+6 \delta_{GX}-6\delta_{GXX}+2 \delta_{G\phi}-2\delta_{PXX}\right)^{1/2}}{8\pi^2 M_{pl}^2 F\left(\delta_{PX}+2 \delta_{F}+4\delta_{GX}+2 \delta_{G\phi}\right)^{3/2}}.
\label{Power_Spectrum_Scalar}
\eea
Since the curvature perturbation becomes constant
for $c_s k<aH$, we may evaluate the power spectrum at the Hubble crossing time, i.e. $c_s k=aH$. Thus, the spectral index of $\mathcal{R}$ is given by
\bea
&& n_{s}-1\equiv \frac{d \ln \mathcal{P}_{s}}{d \ln k}|_{c_{s}k=a H}=-2 \epsilon-\delta_{F}-\eta_{s}-s,\nonumber\\
&& = -2\epsilon_{s}-\eta_{s}-s+\delta_{F}+2 \delta_{GX}+\mathcal{O}(\epsilon^2),
\label{ns}
\eea
where 
\be
\eta_{s}\equiv \frac{\dot{\epsilon}_{s}}{H\epsilon_{s}},\:\:\:\:\: s\equiv \frac{\dot{c}_{s}}{H c_{s}}.
\label{etas}
\ee As usually we have assumed that $c_{s}$ is a slowly varying function, such that $s\ll 1$ \cite{Baumann:2009ds}.

Now, let us calculate the power spectrum of tensor perturbations.
The second-order action for the tensor modes is given by
\be
S_{T}=\sum_{\lambda} \int{dt d^3x a^3 Q_{T}\left[\dot{h}_{\lambda}^2-\frac{c_{T}^2}{a^2} \left(\partial h_{\lambda}\right)^2\right]},
\ee where $\lambda=+,\times$, and we find
\bea
\label{QT}
&& Q_{T}= \dfrac{M^{2}_{pl}}{4} F,\\
\label{CT}
&& c_{T}^2=1.
\eea From Eq. \eqref{QT} one may see that the no-ghost condition $Q_{T}>0$ is satisfied only in the case when $F>0$. On the other hand, Eq. \eqref{CT} guarantees us that there are no Laplacian instabilities for tensor perturbations and also that in our generalized scalar-torsion gravity scenario \eqref{action1}, GWs propagate exactly at the speed of light provided that the coupling function only depends on the scalar field, i.e. $F=F(\phi)$. For more details, see section \ref{Appendix}.
Here, it is also important to point up that the second-order action for the tensor modes is invariant under a Lorentz rotation of the frame $e^{A}_{~\mu}$, and therefore there is no contributions coming from additional degrees of freedom.

Thus, the power spectrum of tensor perturbations is
\be
\mathcal{P}_{T}=\frac{H^2}{2 \pi^2 Q_T}=\frac{2 H^2}{\pi^2 M_{pl}^2 F},
\ee
and the corresponding spectral index is given by
\bea
&& n_{T}\equiv \frac{d \ln \mathcal{P}_{T}}{d \ln k}|_{c_{s}k=a H}=-2 \epsilon-\delta_{F},\nonumber\\
&& =-2 \epsilon_{s}+\delta_{F}+2 \delta_{GX}+\mathcal{O}(\epsilon^2).
\eea
In the phase before the end of inflation when $\mathcal{P}_{s}$ and $\mathcal{P}_{T}$ remain approximately constant, we may evaluate the tensor-to-scalar ratio as
\be
r = \dfrac{\mathcal{P}_{T}}{\mathcal{P}_{s}}= 16 c_s \epsilon_s.
\label{r1}
\ee The above expression can be contrasted with the following consistency relation 
\be
r\simeq 8 c_{s} \left[-n_{T}+\delta_{F}+2 \delta_{GX}\right].
\label{r2}
\ee In standard inflation one has that $r\simeq -8 c_{s} n_{T}$, and therefore $\delta_{F}$ and $\delta_{GX}$ induce small deviations from those relation derived in standard inflation.

\section{Theories with $P=-X+V, G=0, F\neq 0$}{\label{Coupling}}

\subsection{Slow-roll analysis}

For our first particular case, we focus on a model in which inflation is mainly driven by
both a field potential, $P=-X+V$ and a non-minimal coupling $F\neq 0$, with $G=0$.
For this case the background equations \eqref{00} and \eqref{phi} take the following form
\bea
&& 3 H^2 M_{pl}^2 F = \dfrac{1}{2} \dot{\phi}^2 + V,\\
&& \ddot{\phi} + 3 H \dot{\phi} + 3 H^2 M_{pl}^2 F_{,\phi} + V_{,\phi} = 0.
\eea

These equations are exactly the same background equations (10), (11) and (12) (only two of them are independent) of Ref. \cite{Kanno:2006ty}. Therefore, in the FRW background, this non-minimally coupled scalar-torsion theory is equivalent to a scalar-vector-tensor theory, with spontaneous violation of particle local Lorentz invariance due to a time-like vector field. In the spacetime, the torsion tensor can be decomposed into three components, irreducible under the global Lorentz group; there will be a vector field $\mathcal{V}_{\mu}=T^{\nu}_{~\nu \mu}$, an axial part $\mathcal{A}^{\mu}=(1/6) \epsilon^{\mu\nu\rho\sigma} T_{\nu\rho\sigma}$, and a purely tensor part $\mathcal{T}_{\lambda \nu\mu}=(1/2)\left( T_{\lambda \mu\nu}+T_{\mu \lambda \nu}\right)+(1/6)\left(g_{\nu\lambda}\mathcal{V}_{\mu}+g_{\nu\mu}\mathcal{V}_{\lambda}\right)-(1/3) g_{\lambda\mu}\mathcal{V}_{\nu}$ \cite{Aldrovandi-Pereira-book}. So, in the FRW background \eqref{veirbFRW}, the only contribution to the torsion scalar, $T=T^{\rho}_{~\mu\nu}S_{\rho}^{~\mu\nu}$, comes from the vector torsion $\mathcal{V}_{\mu}=(-3 H, 0,0,0)$, which defines a preferred frame as due to a time-like vector field. Thus, once that the non-minimal coupling function is turned on, the particle local Lorentz invariance is spontaneously broken due to the presence of this vector field with vacuum expectation value $\bra{0}\mathcal{V}_{\mu}\mathcal{V}^{\mu}\ket{0}=-9 H^2$, with $H=\sqrt{\lambda/(3 \xi M_{pl}^2)}$ (see also Refs. \cite{Jacobson:2000xp,Carroll:2004ai}).

In the slow-roll approximation $\dot{\phi}^2\ll V$, $|\ddot{\phi}|\ll | H \dot{\phi}|$, these equations acquire the form
\bea
\label{slow1}
&&3 H^2 M_{pl}^2 F\simeq V,\\
&& 3 H \dot{\phi}+3 H^2 M_{pl}^2 F_{,\phi} +V_{,\phi}\simeq 0. 
\label{slow2}
\eea 
Thus, from equation \eqref{slow1} we obtain 
\be
3 H^2\simeq \frac{V}{M_{pl}^2 F},
\ee whereas that from equation \eqref{slow1} and \eqref{slow2} we obtain
\be
\frac{\dot{\phi}}{M_{pl} H}\simeq -\left[\left(\frac{M_{pl} V_{,\phi}}{V}\right)+\left(\frac{M_{pl} F_{,\phi}}{F}\right)\right] F.
\label{slow3}
\ee Thus, for $F>0$, $V_{,\phi}>0$ and $F_{,\phi}>0$, one has $\dot{\phi}<0$, whereas that for $F>0$, $V_{,\phi}<0$ and  $F_{,\phi}<0$, it is satisfied  $\dot{\phi}>0$.

Now, by using the equation \eqref{SlowRoll-Parameters} for the slow-roll parameters $\delta_{F}$ and $\delta_{PX}$, it is straightforward to see that 
\be
\delta_{F}=\frac{\dot{F}}{H F}=\left(\frac{M_{pl} F_{,\phi}}{F}\right)\left(\frac{\dot{\phi}}{M_{pl} H}\right),
\label{deltaF2}
\ee and
\be
\delta_{PX}=\frac{-X P_{,X}}{M_{pl}^2 H^2 F}=\left(\frac{1}{2 F}\right) \left(\frac{\dot{\phi}}{M_{pl} H}\right)^2,
\label{deltaPX2}
\ee and thus, from Eq. \eqref{epsilon2} one finds 
\be
\epsilon\simeq \left(\frac{M_{pl} F_{,\phi}}{F}\right)\left(\frac{\dot{\phi}}{M_{pl} H}\right)+\left(\frac{1}{2 F}\right) \left(\frac{\dot{\phi}}{M_{pl} H}\right)^2.
\ee By substituting \eqref{slow3} into this latter equation gives us
\be
\epsilon\simeq \left(\epsilon_{V}-\epsilon_{F}\right) F,
\label{epsilonF1}
\ee where we have defined the non-minimal coupling slow-roll parameter $\epsilon_{F}$ and the potential slow-roll parameter $ \epsilon_{V}$ as 
\be
\epsilon_{F}\equiv \frac{1}{2}\left(\frac{M_{pl} F_{,\phi}}{F}\right)^2,\:\:\:\:\: \epsilon_{V}\equiv \frac{1}{2}\left(\frac{M_{pl} V_{,\phi}}{V}\right)^2.
\label{epF_ep_V}
\ee 
Under the slow-roll approximation, the number of $e$-folds $N\equiv \log(a)$ measuring the amount of inflation between the time
around the cosmological scales cross the Hubble radius, $t_{*}$, and the end of inflation $t_f$, is calculated as 
\be
N\equiv \int_{t_{*}}^{t_{f}}{H dt}\simeq \int_{\phi_{f}}^{\phi_{*}}{\frac{1}{\left[\left(\frac{M_{pl} V_{,\phi}}{V}\right)+\left(\frac{M_{pl} F_{,\phi}}{F}\right)\right]F} \frac{d\phi}{M_{pl}}}.
\label{Nefolds}
\ee In the above equation the value of the field at the end of inflation, $\phi_{f}$, is calculated from the condition $\epsilon(\phi_{f})\simeq 1$.

By using Eqs. \eqref{slow3}, \eqref{deltaF2}, and \eqref{deltaPX2}, the parameters $Q_{s}$ and $c_{s}^2$ in Eqs. \eqref{Qs} and \eqref{csSlow}, respectively, are written as 
\bea
\label{Qs1}
&& Q_{s}\simeq M_{pl}^2 F \delta_{PX}>0, \\ 
&& c_{s}^2\simeq 1+2 \frac{\delta_{F}}{\delta_{PX}}.
\label{csSlow1}
\eea From Eq. \eqref{csSlow1}, it can be seen that $c_{s}^2> 0$  for $\delta_{F}/\delta_{PX}> -1/2$.

In a similar way, the power spectra of scalar perturbations (Eq. \eqref{Power_Spectrum_Scalar}) becomes
\bea
&& \mathcal{P}_{s}\simeq \left[\frac{V}{ 24 \pi^2 M_{pl}^4 F^{2}}\right]\frac{\left(\delta_{PX}\right)^{1/2}}{\left(\delta_{PX}+2 \delta_{F}\right)^{3/2}},
\label{PowerSpectrum1}
\eea where we have also used the slow-roll equation \eqref{slow1}.

On the other hand, we can also obtain the corresponding expressions for the scalar spectral $n_s $ and the tensor-to-scalar ratio $r$, when we are considering a non-zero coupling function $F$. In doing so, we replace Eqs.
\eqref{deltaF2}, \eqref{deltaPX2},\eqref{epsilonF1},  \eqref{epF_ep_V}, and \eqref{csSlow1},
in Eqs. \eqref{ns} and \eqref{r2}, respectively (not shown).

In order to obtain concrete results, we are going to consider a specific expression for both the non-minimal coupling function $F$ and the inflaton potential $V(\phi)$.

\subsection{Chaotic Inflation}

Let us consider that the inflaton potential has a power-law form
\be
V(x)=\frac{\lambda}{n} x^{n},
\label{PowerLawPotential}
\ee where $\lambda$, $n$ are positive constants and  $x\equiv \phi/M_{pl}$. In this case one obtains that after replacing \eqref{PowerLawPotential} in Eq. \eqref{epF_ep_V}, $\epsilon_V$ becomes
\be
\epsilon_{V}=\frac{n^2}{2}x^{-2}.
\label{epsilonV2}
\ee Similarly, for the non-minimal coupling function let us assume the ansatz
\be
F(x)=1+\frac{\xi}{2} x^{2},
\label{PowerLawNMF}
\ee where $\xi$ is a positive constant, which guarantees us the no-ghost condition $F>0$. Then, by replacing Eq. \eqref{PowerLawNMF} in Eq. \eqref{epF_ep_V}, the slow-roll parameter $\epsilon_{F}$ has the following dependence on the scalar field
\be
\epsilon_{F}(x)=\frac{1}{2}\left[\frac{\xi x}{1+\frac{\xi}{2}x^{2}}\right]^{2}.
\label{Chao_epF}
\ee
Accordingly, by combining Eq. \eqref{epsilonF1} with Eqs. \eqref{epsilonV2}-\eqref{Chao_epF}, it is straightforward to obtain
\be
\epsilon(x)= \frac{n^2 \left[\frac{\xi  x^2}{2}+1\right] \left[1-\frac{\xi ^2 x^{4}}{n^2 \left(\frac{\xi  x^2}{2}+1\right)^2}\right]}{2 x^2}.
\label{eps_Coup}
\ee

Inflation ends at $\epsilon(x_{f})=1$, from which we obtain the value of
the inflaton field at the end of inflation $x_f$. Therefore, from Eq. \eqref{eps_Coup} one finds
\be
x_{f} = \sqrt{\frac{n^2}{1-\frac{\xi  n^2}{2}+\sqrt{n^2 \xi ^2+1}}}.
\label{phif_Coupl}
\ee

Now, from Eq. \eqref{Nefolds} we compute the number of $e$-folds $N$ as
\bea
&& N= \int_{t*}^{t_{f}}{H dt}\simeq \int_{x_{f}}^{x_{*}}{\frac{\frac{x}{n}dx}{\left[1+\left(\frac{1}{m}+\frac{1}{n}\right)\tilde{\xi}x^{m}\right]}},\nonumber\\
&& =\frac{\log \left[(n+2) \xi  x_{*}^2+2 n\right]-\log \left[\frac{4 n \left(\sqrt{n^2 \xi ^2+1}+n \xi +1\right)}{2 \sqrt{n^2 \xi ^2+1}-\xi  n^2+2}\right]}{(n+2) \xi }.
\label{e_folds_N}
\eea

\begin{figure}[htbp]
\begin{center}
\includegraphics[width=0.6\textwidth]{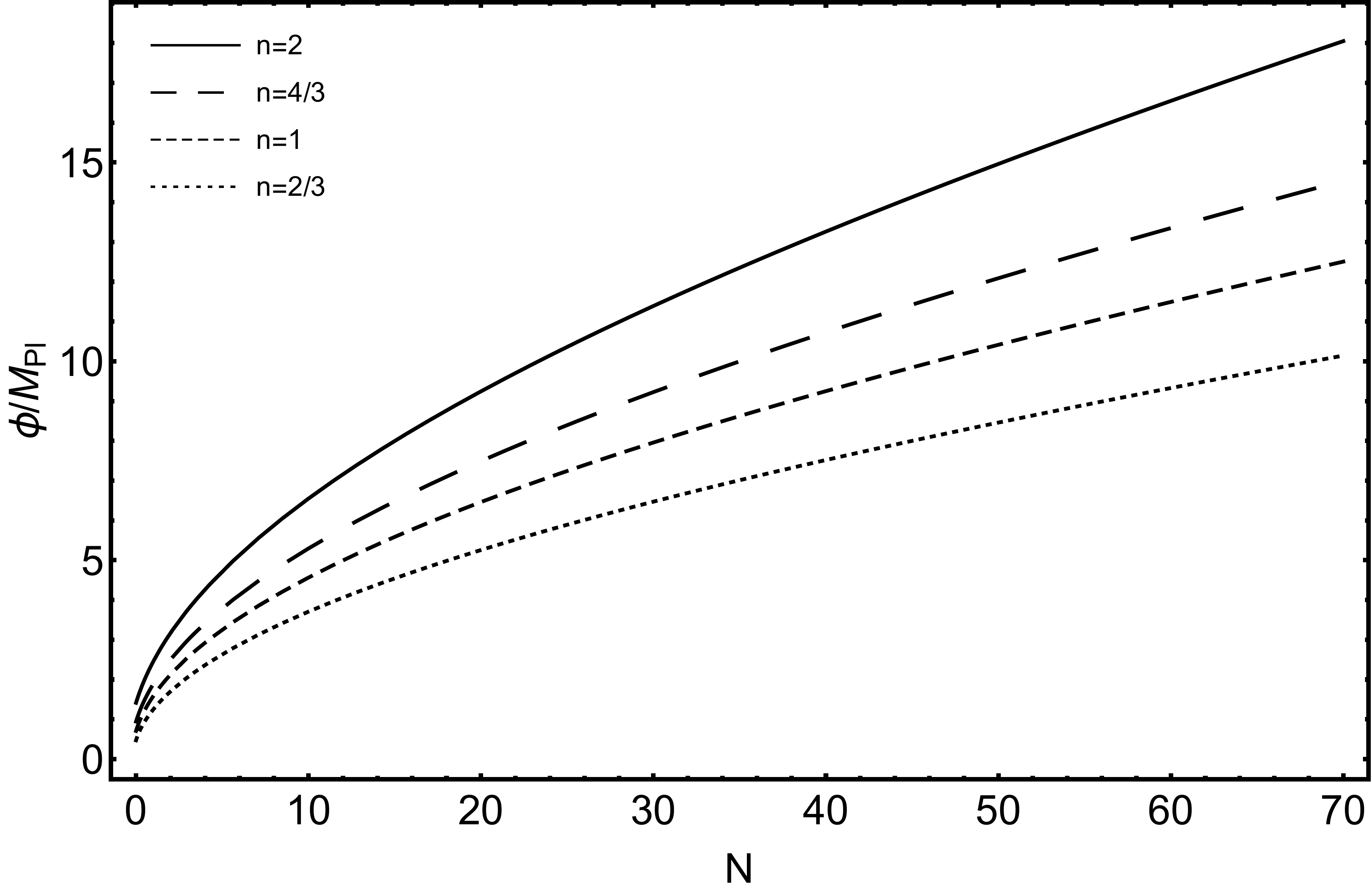}
\includegraphics[width=0.6\textwidth]{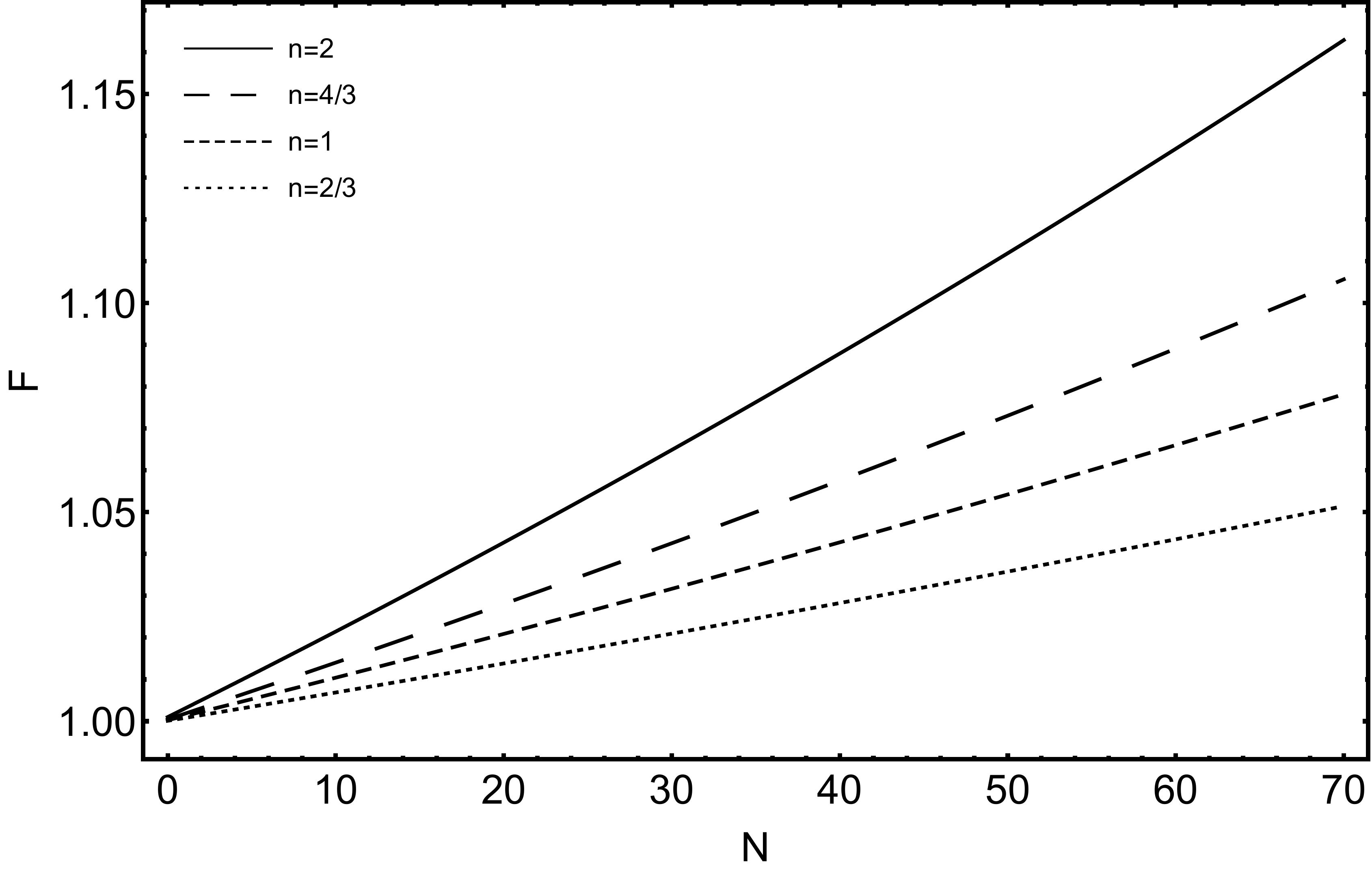}
\caption{\it{Evolution of the scalar field $x\equiv \phi/M_{pl}$ (upper graph) and the non-minimal coupling function $F(x)=1+\xi x^2/2$ (lower graph), with coupling $\xi=0.001$, and for the power-law potential $V(x)=\frac{\lambda}{n} x^{n}$ in four different cases of
values for the power $n$.}}
\label{x_Coupling}
\end{center}
\end{figure}

This latter equation can be solved for $x_{*}$, such that
\bea
x_{*}=\sqrt{e^{(n+2) \xi N} \left[\frac{2 n}{(n+2) \xi }+ x_{f}^2\right]-\frac{2 n}{(n+2) \xi }},
\label{phi_N_Coup} 
\eea where $x_{f}$ is given in Eq. \eqref{phif_Coupl}. Therefore, from this equation we get the value of the scalar field at the time of Hubble-radius crossing, i.e. $x_{*}$, expressed
as a functions of the number of $e$-folds $N$, the power $n$ and the coupling constant $\xi$. In FIG \ref{x_Coupling} (upper graph) we plot the behaviour of the scalar field $x_{*}$ against the number of $e$-folds $N$, for a specific value of the non-minimal coupling  $\xi=0.001$ and some special values of the power $n$ \cite{Akrami:2018odb}. Also, in FIG \ref{x_Coupling} (lower graph) we show the evolution of the non-minimal coupling function $F$ against the number of $e$-folds, with the same values for $\xi$ and $n$.  As we can see, the field takes values above the Planck scale, i.e. $x>1$ at the horizon crossing, and that $F$ evolves in an intermediate regime between $F-1=\frac{\xi}{2} x^2\ll 1$ and $F-1=\frac{\xi}{2} x^2\gg 1$, with $F\sim 1$.

Additionally, from Eqs. \eqref{deltaF2} and \eqref{deltaPX2} one finds
\be
\delta_{F}=-\xi \left[n+\frac{2\xi x^2}{2+\xi  x^2}\right],
\label{deltaF_Coupl}
\ee and
\bea
&& \delta_{PX}=\frac{\left[(n+2) \xi  x^2+2 n\right]^2}{4 x^2 \left(\xi  x^2+2\right)}.
\label{deltaPX_Coupl}
\eea

\begin{figure}[htbp]
\begin{center}
\includegraphics[width=0.6\textwidth]{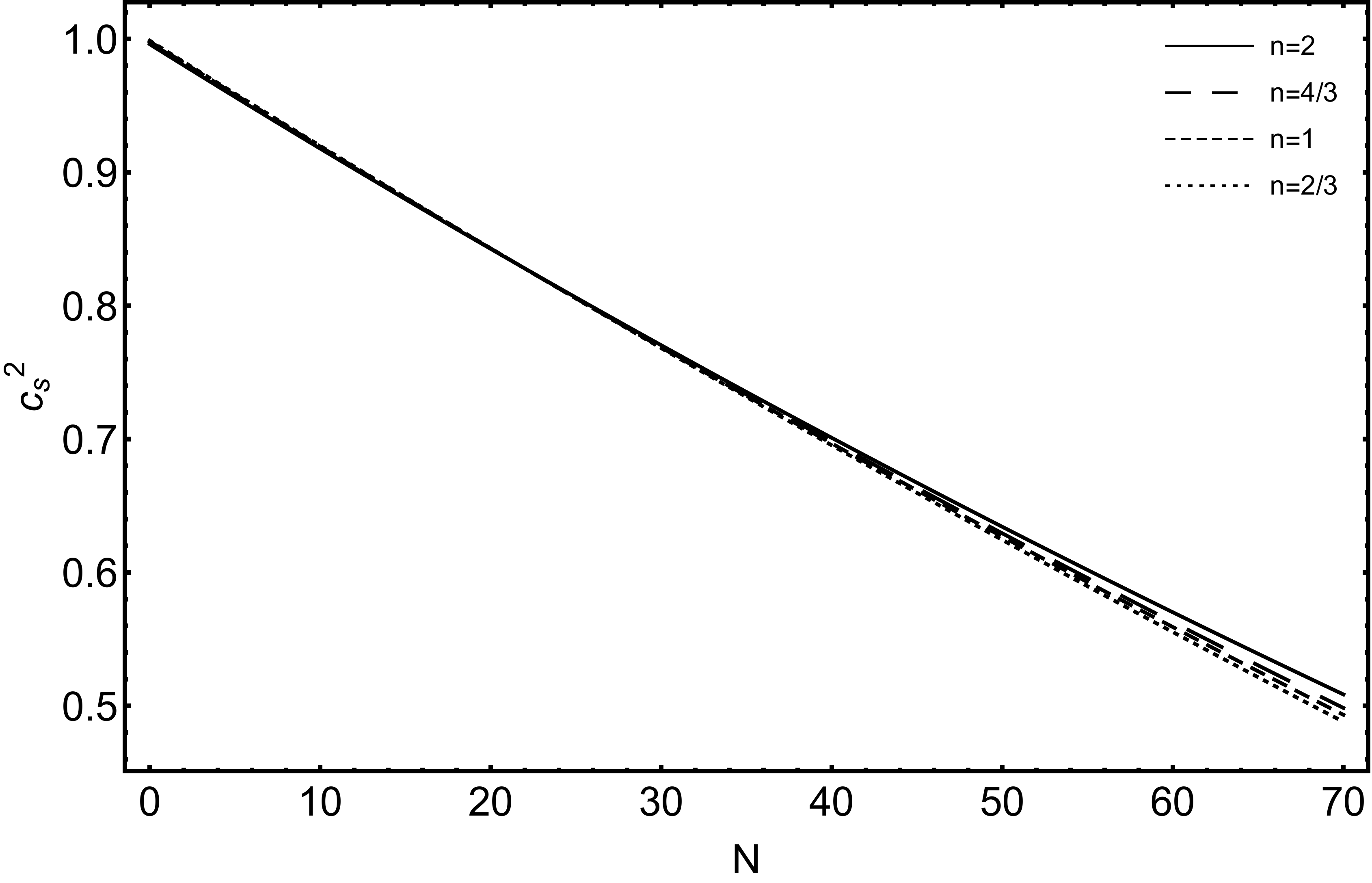}
\caption{\it{The evolution of scalar propagation speed squared $c_{s}^2$ in the presence of the non-minimal coupling function $F(x)=1+\xi x^2/2$ with coupling $\xi=0.001$, and for the chaotic potential $V(x)=\frac{\lambda}{n} x^{n}$, in the case of four different values of $n:2,4/3,1,2/3$. }}
\label{cs_Coupling}
\end{center}
\end{figure}

Thus, using the above equations, \eqref{deltaF_Coupl} and \eqref{deltaPX_Coupl}, into Eqs. \eqref{Qs1} and \eqref{csSlow1}, for the $Q_{s}$ parameter one obtains
\bea
Q_{s}= \frac{M_{pl}^2 \left[(n+2) \xi  x^2+2 n\right]^2}{8 x^2}\geq 0, 
\eea whereas the field dependence of the scalar propagation speed squared is found to be
\be
c_{s}^2= 1-\frac{8 \xi  x^2}{(n+2) \xi  x^2+2 n}.
\label{cs_Coup}
\ee  In FIG \ref{cs_Coupling} we show the evolution of $c_{s}^2$ as function of the $e$-folds number $N$ for fixed value $\xi=0.001$ and several values of the power-law exponent $n$. The requirement $c_{s}^2>0$ is satisfied for $0<x<\sqrt{\frac{2 n}{\left(6-n\right) \xi}}$, with $0<n\leq 4$ and $\xi>0$.

The scalar power spectrum is calculated by using \eqref{PowerSpectrum1}, along with Eqs. \eqref{deltaF_Coupl} and \eqref{deltaPX_Coupl}, which yields
\bea
&& \mathcal{P}_{s}=\frac{2 \tilde{\lambda}  x^{n+2} \left(\xi  x^2+2\right)^{-1}}{3 \pi ^2 n \sqrt{\left[(n-6) \xi  x^2+2 n\right]^{3}\left[(n+2) \xi  x^2+2 n\right]}}
\label{PN},
\eea where we have also defined the dimensionless parameter $\tilde{\lambda}\equiv\lambda/M_{pl}^4$ and $x$ is given in Eq. \eqref{phi_N_Coup}. 

\begin{figure}[htbp]
\begin{center}
\includegraphics[width=0.6\textwidth]{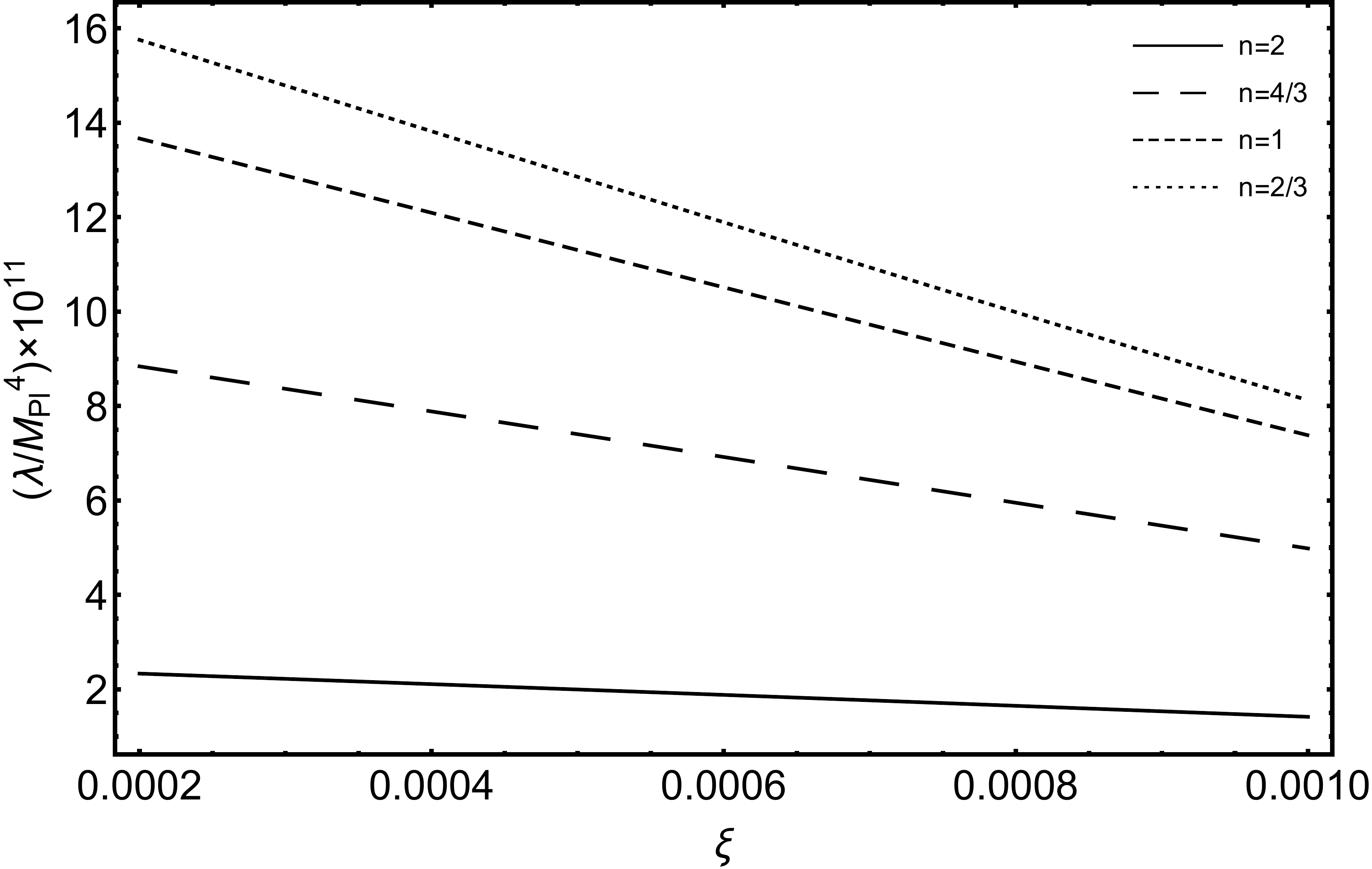}
\caption{\it{We plot the relation $\lambda(\xi)$ for the theories $P=-X+V, G=0, F\neq0$, with $V(x)=\frac{\lambda}{n} x^{n}$ and $F(x)=1+\xi x^2/2$, for $\mathcal{P}_{s}\simeq 2.141\times 10^{-9}$ at $N=70$, and several values of the power $n$.}}
\label{FIG_lambda_Coupling}
\end{center}
\end{figure}

After evaluating \eqref{PN} at the value of the scalar field when a given perturbation scale leaves the Hubble-radius, given by \eqref{phi_N_Coup}, and by using the current observational value for the amplitude of primordial scalar perturbations $\mathcal{P}_{s}= 2.141 \times 10^{-9}$ \cite{Ade:2015lrj,Akrami:2018odb}, we may find a constraint which relates $\tilde{\lambda}$ to $n$, $\xi$, and $N$. In FIG \ref{FIG_lambda_Coupling} we show the behaviour of $\lambda$ in terms of $\xi$ for several values of the power $n$ when the number of $e$-folds is fixed to be $N=70$. For all the cases, it can be seen that $\lambda$ is a  monotonically decreasing function with the increasing of the non-minimal coupling parameter $\xi$.

In order to calculate the spectral index and the tensor-to-scalar ratio
we obtain at first the slow-roll parameter $\epsilon_{s}$. By putting Eqs. \eqref{eps_Coup} and \eqref{deltaF_Coupl} into Eq. \eqref{epsilons2} one gets
\be
 \epsilon_{s}\simeq \epsilon+\delta_{F}= \frac{\left[(n-6) \xi  x^2+2 n\right] \left[(n+2) \xi  x^2+2 n\right]}{4 x^2 \left(\xi  x^2+2\right)}.
\label{epss_Coup}
\ee

Therefore the slow-roll parameter $\eta_{s}$ (Eq. \eqref{etas}) yields
\be
 \eta_{s}=\frac{\dot{\epsilon}_{s}}{H\epsilon_{s}}= -\frac{4 \xi }{\xi  x^2+2}+ 2 n \left[\frac{8 \xi }{(n-6) \xi  x^2+2 n}+\frac{1}{x^2}\right].
\label{etas_Coup}
\ee Moreover, from Eqs. \eqref{etas} and  \eqref{cs_Coup}, we calculate the slow-roll parameter $s$ which becomes
\be
s=\frac{\dot{c}_{s}}{H c_{s}}=\frac{8 n \xi }{(n-6) \xi  x^2+2 n}.
\label{s_Coupling}
\ee

In this form, the scalar spectral index of $\mathcal{R}$ as a function of 
the scalar field is calculated by putting Eqs. \eqref{eps_Coup}, \eqref{deltaF_Coupl}, \eqref{etas_Coup} and \eqref{s_Coupling}, into Eq. \eqref{ns}, which yields
\bea
&& n_{s}= 1-\frac{n (n+2)}{x^2}+\xi  \left[-\frac{n^2}{2}-\frac{24 n}{(n-6) \xi  x^2+2 n}+n-\frac{4}{\xi  x^2+2}+4\right]\label{nsf}.
\eea 

On the other hand, the scalar field dependence of the tensor-to-scalar ratio is calculated from Eq. \eqref{r1}, along with the equations \eqref{cs_Coup} and \eqref{epss_Coup}, giving
\bea
&& r=\left[\frac{4 \left((n-6) \xi  x^2+2 n\right) \left((n+2) \xi  x^2+2 n\right)}{x^2 \left(\xi  x^2+2\right)}\right]\sqrt{1-\frac{8 \xi  x^2}{(n+2) \xi  x^2+2 n}}.\label{rf}
\eea 

After evaluating these inflationary observables at the value of the scalar field when a given perturbation scale leaves the Hubble-radius, given by \eqref{phi_N_Coup}, we may compare the theoretical predictions of this particular model in the $n_s-r$ with the allowed contour regions from Planck 2018 data. Then, we find the allowed ranges of the parameters that characterize this subclass of model. 

\begin{figure}[htbp]
\begin{center}
\includegraphics[width=0.8\textwidth]{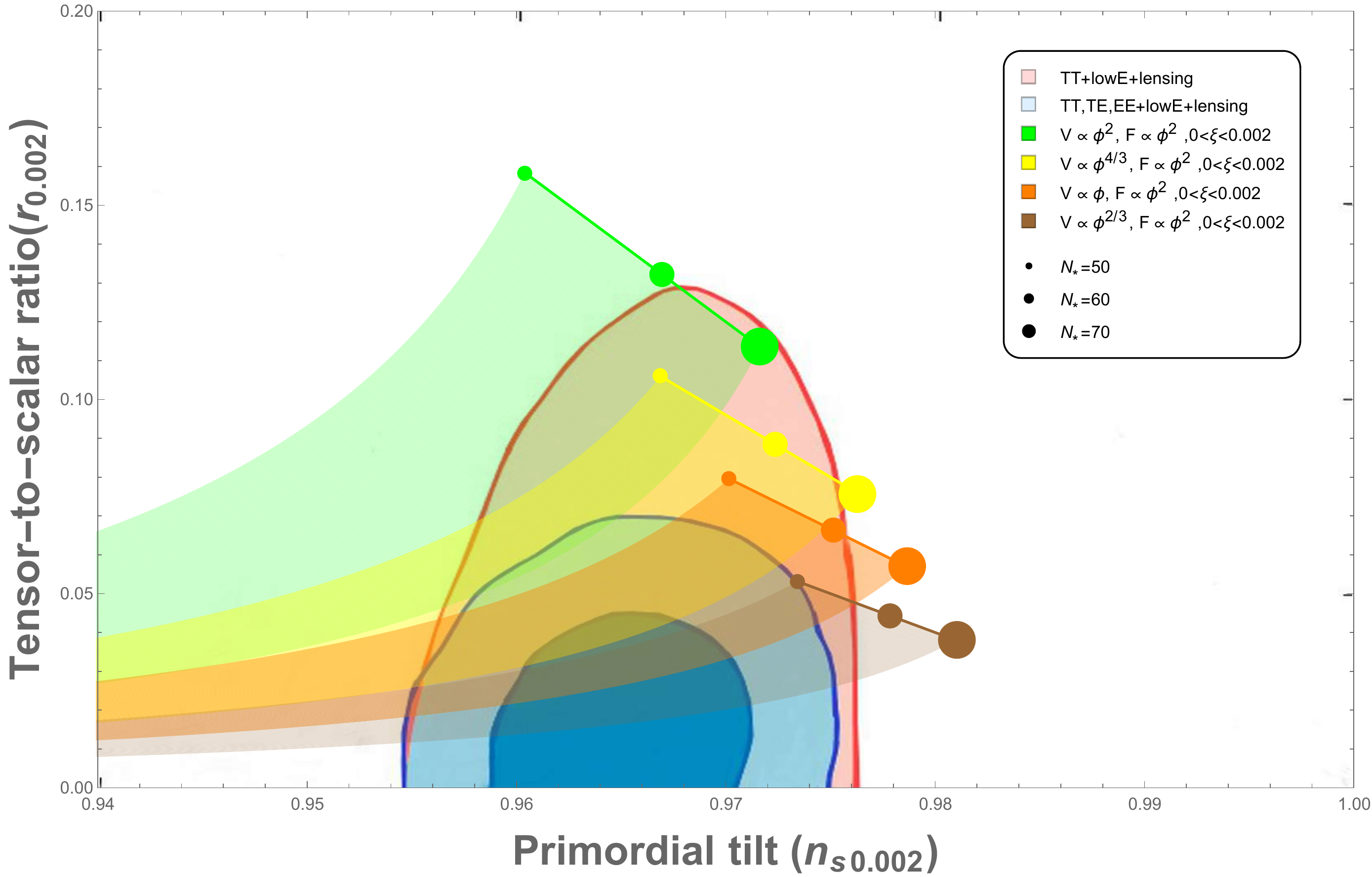}
\caption{\it{We show the plot of the tensor-to-scalar ratio $r$ versus the scalar spectral index $n_s$ for theories with $P=-X+V, G=0, F=1+\xi x^2/2$ along with the two-dimensional
marginalized joint confidence contours for $(n_s,r)$, at the 68$\%$ and 95$\%$ C.L., from
the latest Planck $2018$ results \cite{Akrami:2018odb}, for several values of the power $n:2, 4/3, 1, 2/3$. The quadratic inflation scenario ($n=2$) is compatible with current data at least in the 95$\%$ C.L., whereas that for $n<2$ the predictions of the model are within the contour at 68$\%$ C.L.}}
\label{FIG r_ns_Coupling}
\end{center}
\end{figure}

The trajectories in the $n_s-r$ plane for the model studied here may be generated by plotting Eqs. \eqref{nsf} and \eqref{rf} (both evaluated at $x_{*}$) parametrically, varying both the number of $e$-folds $N$ and the coupling parameter $\xi$ in a wide range. FIG \ref{FIG r_ns_Coupling} shows the plot of the tensor-to-scalar ratio $r$ versus the scalar
spectral index $n_s$  corresponding to this class of non-minimal coupling model satisfying the relations \eqref{PowerLawPotential} and \eqref{PowerLawNMF}, for several values of the power $n$. Here we have considered the two-dimensional marginalized joint confidence contours for $(n_s,r)$, at
the $68\%$ and $95\%$ C.L., from the latest Planck data \cite{Ade:2015lrj,Akrami:2018odb}.
In the particular case of quadratic inflation scenario  $n=2$, one obtains that 
only at $N=70$, the predictions of the model are within the $95\%$ C.L. region from Planck data \cite{Ade:2015lrj,Akrami:2018odb}, for $\xi$ being within the range
\be 
7.10\times 10^{-4} \lesssim \xi_{n=2,N=70} \lesssim 1.08 \times 10^{-3}\label{R_xi_1}.
\ee In that case, the prediction for the tensor-to-scalar ratio is 
$0.05\lesssim r\lesssim 0.07$.

On the other hand, for $0<n<2$, the prediction of model are within the $68\%$ C.L.  for the following ranges of $\xi$
\bea
\label{R_xi_2}
&& 4.20 \times 10^{-4}\lesssim \xi_{n=\frac{4}{3},N=70} \lesssim 1.23\times 10^{-3},\\
\label{R_xi_3}
&& 2.90\times 10^{-4} \lesssim \xi_{n=1,N=60} \lesssim 1.25\times 10^{-3},\\
&&  3.20 \times 10^{-4}\lesssim \xi_{n=\frac{2}{3},N=60} \lesssim 1.29\times 10^{-3}.
\label{R_xi_4}
\eea

Thus, by using these previous results, we find that, depending on the value of the power $n$, the corresponding allowed ranges for $\tilde{\lambda}=\lambda/M_{pl}^4$
are found to be
\bea
\label{Scalar_mass}
&& 1.32 \times 10^{-11}\lesssim\tilde{\lambda}_{n=2,N=70}\lesssim 1.75\times 10^{-11},\\
&& 3.89\times 10^{-11}\lesssim\tilde{\lambda}_{n=\frac{4}{3},N=70}\lesssim 7.79\times 10^{-11},\\
&& 8.60 \times 10^{-11}\lesssim\tilde{\lambda}_{n=1,N=60}\lesssim 1.67\times 10^{-10},\\
&& 8.78 \times 10^{-11} \lesssim\tilde{\lambda}_{n=\frac{2}{3},N=60}\lesssim 1.85  \times 10^{-10}.
\eea 
From the first constraint in Eq. \eqref{Scalar_mass}, we may infer
the allowed values for the mass of the inflaton field, $m_{\phi}=\sqrt{\tilde{\lambda}} M_{pl}$, for quadratic chaotic inflation and $N=70$, are given by,
\bea
&& 3.64\times 10^{-6} \lesssim m_{\phi}/M_{pl} \lesssim 4.19\times 10^{-6}.
\eea 
For the special case  of the chaotic quadratic potential $n=2$, we observe that the predicted value for the mass of the inflaton field becomes of the same order than those obtained for the same potential in the standard scenario ($m_{\phi}\sim 10^{-6}$). Nevertheless,
the predictions for this non-minimally coupled scenario regarding the scalar spectral index and the tensor-to-scalar ratio, through
the $n_s-r$ plane, deviate from the standard scenario. Specifically, this model predicts a smaller tensor-to-scalar ratio 
compared to those obtained in the standard scenario ($\xi=0$), bringing
the quadratic chaotic potential compatible with current observations only at the level 
of $95\%$ C.L. region. In addition, this consistency requires a number of $e$-folds greater than $60$. For a sake of comparison with recent works in the literature, in
\cite{Tenkanen:2017jih} (see also Ref. \cite{Gomes:2016cwj}) it was studied quadratic inflation where the inflaton is
non-minimally coupled to the curvature scalar $R$. Interestingly, it was found that
in order to rescue quadratic inflation, the non-minimal coupling $\xi$ must take
values around $\xi_{R} \sim 10^{-3}$ for $N=60$ . Accordingly, the predicted value for the tensor-to-scalar ratio is $0.01 \lesssim r < 0.12$. Therefore, recalling that our particular subclass of
model predicts $0.05\lesssim r \lesssim 0.07$, it follows that it is possible to distinguish quadratic inflation with the inflaton non-minimally coupled to the curvature scalar $R$ from quadratic inflation in a scalar-torsion gravity. Although our model is in consistency with current bounds set by Planck, forthcoming B-mode polarization experiments such as BICEP3 \cite{Wu:2016hul} or LiteBIRD \cite{Suzuki:2018cuy}, expect to set an upper bound to tensor-to-scalar ratio such as $r\lesssim 0.03$ and $r\lesssim 0.001$, respectively. In view of this,
the new experiments may eventually rule out chaotic quadratic inflation in a scalar-torsion scenario
with a non-minimal coupling to torsion in near future.

\section{Theories with $P=-X+V, G=\gamma X, F\neq 0$}\label{Coupling_Galileon}

\subsection{Slow-roll analysis}

In our second particular model, inflation is now mainly driven by
a field potential, $P=-X+V$, the Galileon self-interaction $G=\gamma X$, and the non-minimal coupling
function $F\neq 0$. Then, the background equations \eqref{00} and \eqref{phi} assume the form
\bea
&& 3 H^2  M_{pl}^2 F = \dfrac{1}{2} \dot{\phi}^2 + V - 3 \gamma H \dot{\phi}^3,\\
&& \ddot{\phi} \left[ 1 - 6 \gamma H \dot{\phi} \right] + 3 H \dot{\phi} \left[ 1 - 3 \gamma H \dot{\phi} - \gamma \dfrac{\dot{H}}{H} \dot{\phi} \right] \nonumber\\
&& + 3 H^2 M_{pl}^2 F_{,\phi} + V_{,\phi} = 0, 
\eea
Under the slow-roll approximation $\dot{\phi}^2\ll V$, $|\ddot{\phi}|\ll | H \dot{\phi}|$, this set of equations leads us to
\bea
\label{slowGF1}
&& 3 H^2  M_{pl}^2 F\simeq V,\\
&& 3 H \dot{\phi} \left[ 1 +\mathcal{A}\right] +3 H^2 M_{pl}^2 F_{,\phi} + V_{,\phi}\simeq 0, 
\label{slowGF2}
\eea where $\mathcal{A}\equiv- 3 \gamma H \dot{\phi}$ is a new function which
takes into account the effect of the Galileon self-interaction in
the dynamics of the inflaton field. From \eqref{slowGF2}, $\mathcal{A}$ may be regarded
as an extra friction term, slowing-down the evolution of $\phi$ relative to
those in standard inflation. Thus, from Eqs. \eqref{slowGF1} and \eqref{slowGF2} we get
\bea
&& 3 H^2 \simeq \frac{V}{ M_{pl}^2 F},\\
&& \frac{\dot{\phi}}{M_{pl} H} \simeq -\left[\frac{F}{ 1 +\mathcal{A}}\right]\left[\left(\frac{M_{pl} V_{,\phi}}{V}\right)+\left(\frac{M_{pl} F_{,\phi}}{F}\right)\right] .
\label{dotPhi_Galileon}
\eea Furthermore, by substituting $\mathcal{A}=- 3 \gamma H \dot{\phi}$ in the latter equation we can solve for $\dot{\phi}$ which gives us
\bea
\label{phi2}
&& \dot{\phi}(\phi)\simeq \frac{1}{6 \gamma H}\left[1-\sqrt{1+4 \gamma V\left[\frac{V_{,\phi}}{V}+\frac{F_{,\phi}}{F}\right]}\right],\\
&& \mathcal{A}(\phi)\simeq\frac{1}{2}\left[ -1+\sqrt{1+4 \gamma V \left[\frac{V_{,\phi}}{V}+\frac{F_{,\phi}}{F}\right]}\right].
\label{A1}
\eea Here we can see that if $\gamma<0$, $V_{,\phi}<0$, and $F_{,\phi}<0$ then $\dot{\phi}>0$ and $\mathcal{A}>0$. On the other hand, in the case with $\gamma>0$, $V_{,\phi}>0$, and $F_{,\phi}>0$, one has that $\dot{\phi}<0$ and $\mathcal{A}>0$. The transition from Galileon driven inflation to standard inflation occurs at the value of the field $\phi_{G}$, which is calculated from the condition $\mathcal{A}(\phi_{G})=1$ \cite{Ohashi:2012wf}. From Eq. \eqref{A1}, the latter condition is translated into the following relation 
\be
\frac{1}{2}\gamma\left[V_{,\phi}(\phi_{G})+\frac{F_{,\phi}(\phi_{G}) V(\phi_{G})}{F(\phi_{G})}\right]=1.
\label{PhiG}
\ee

\begin{figure*}[ht]
\begin{center}
\includegraphics[width=0.28\textwidth]{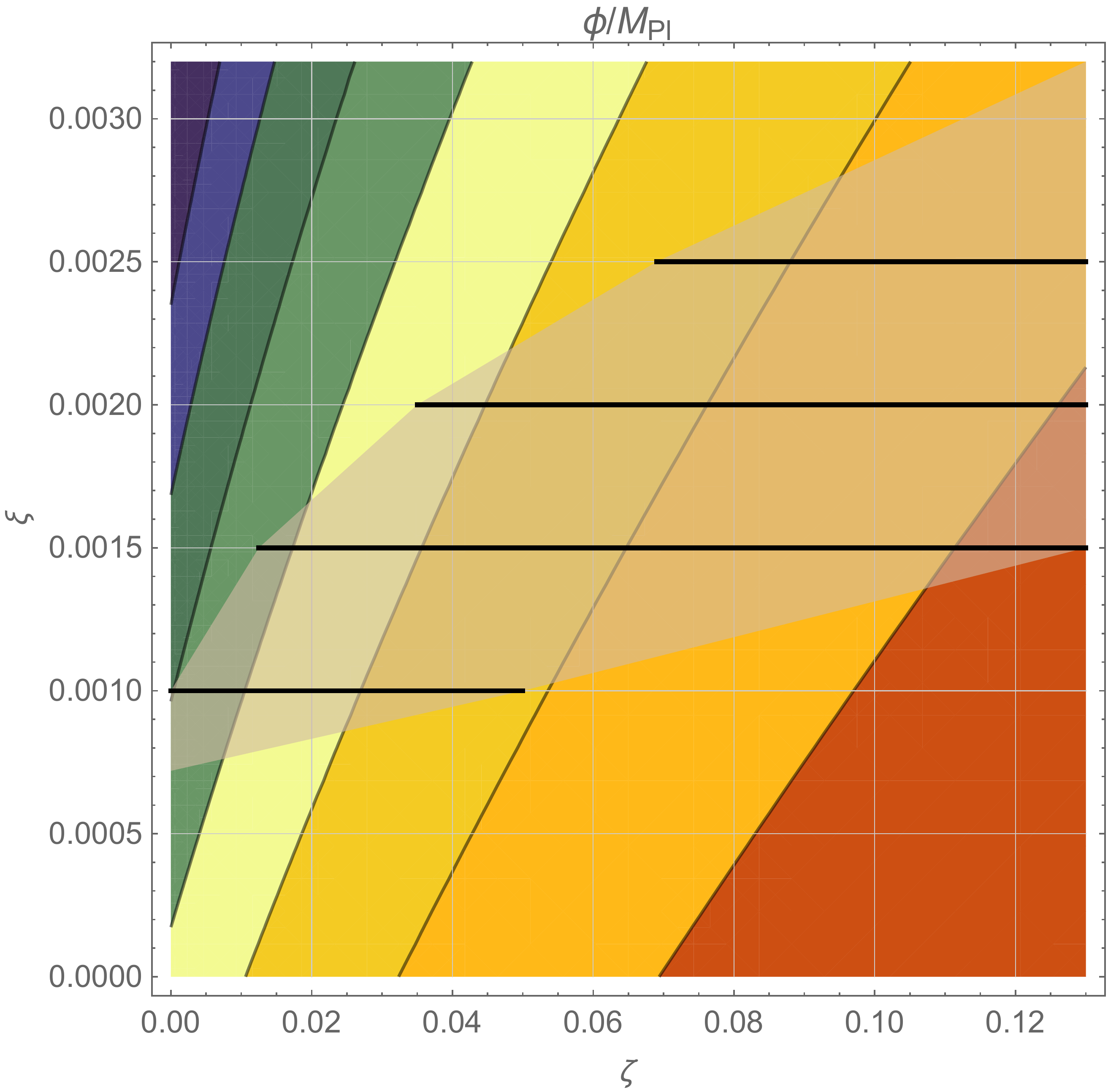}
\includegraphics[width=0.03\textwidth]{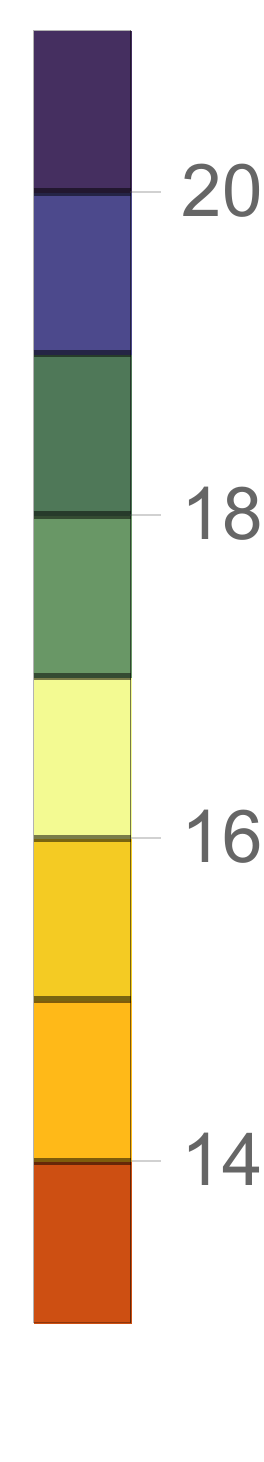}
\includegraphics[width=0.28\textwidth]{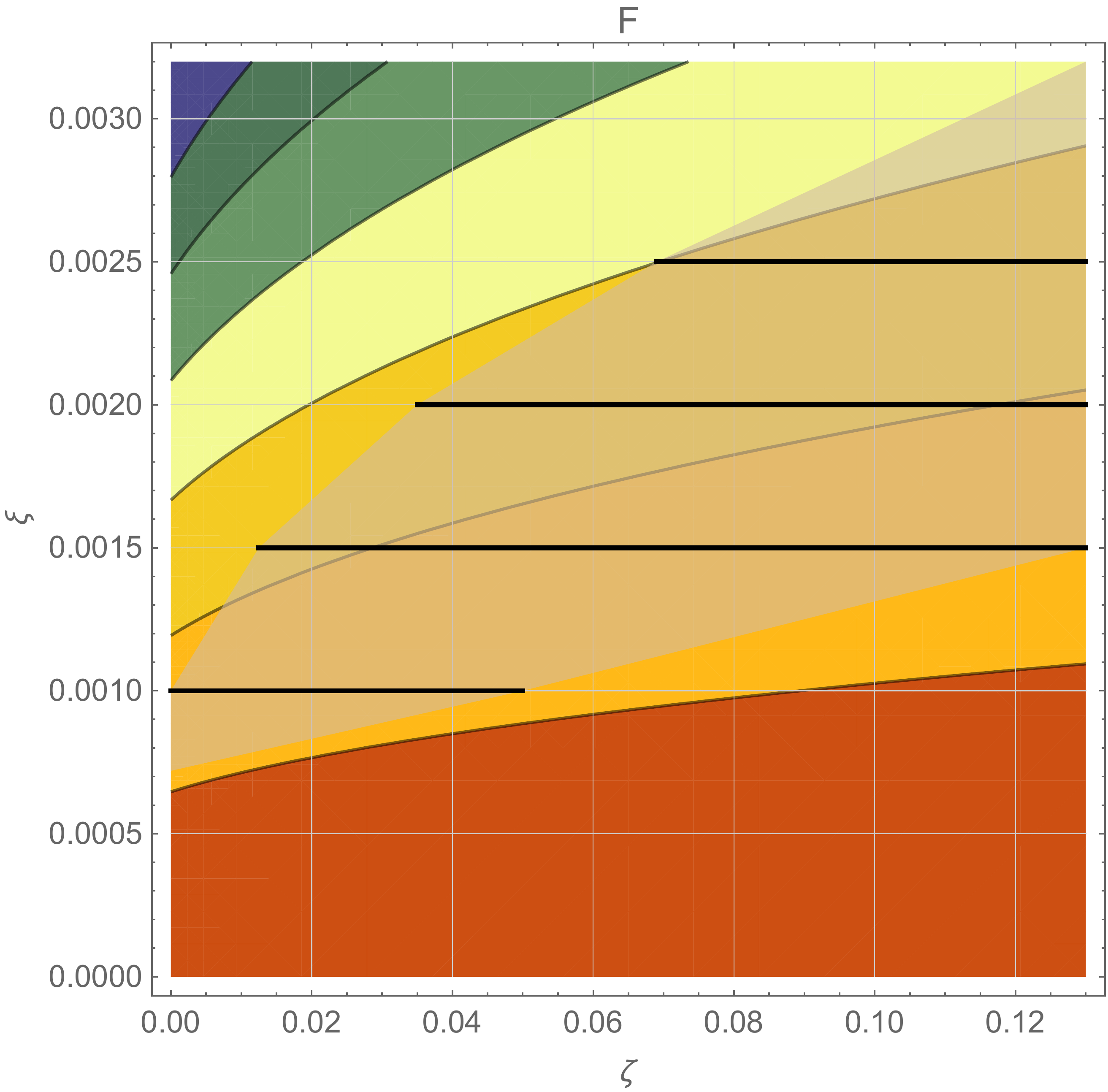}
\includegraphics[width=0.03\textwidth]{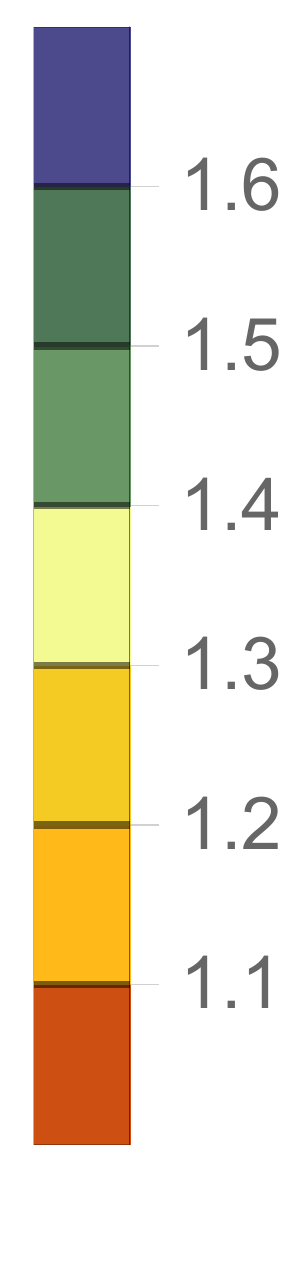}
\includegraphics[width=0.28\textwidth]{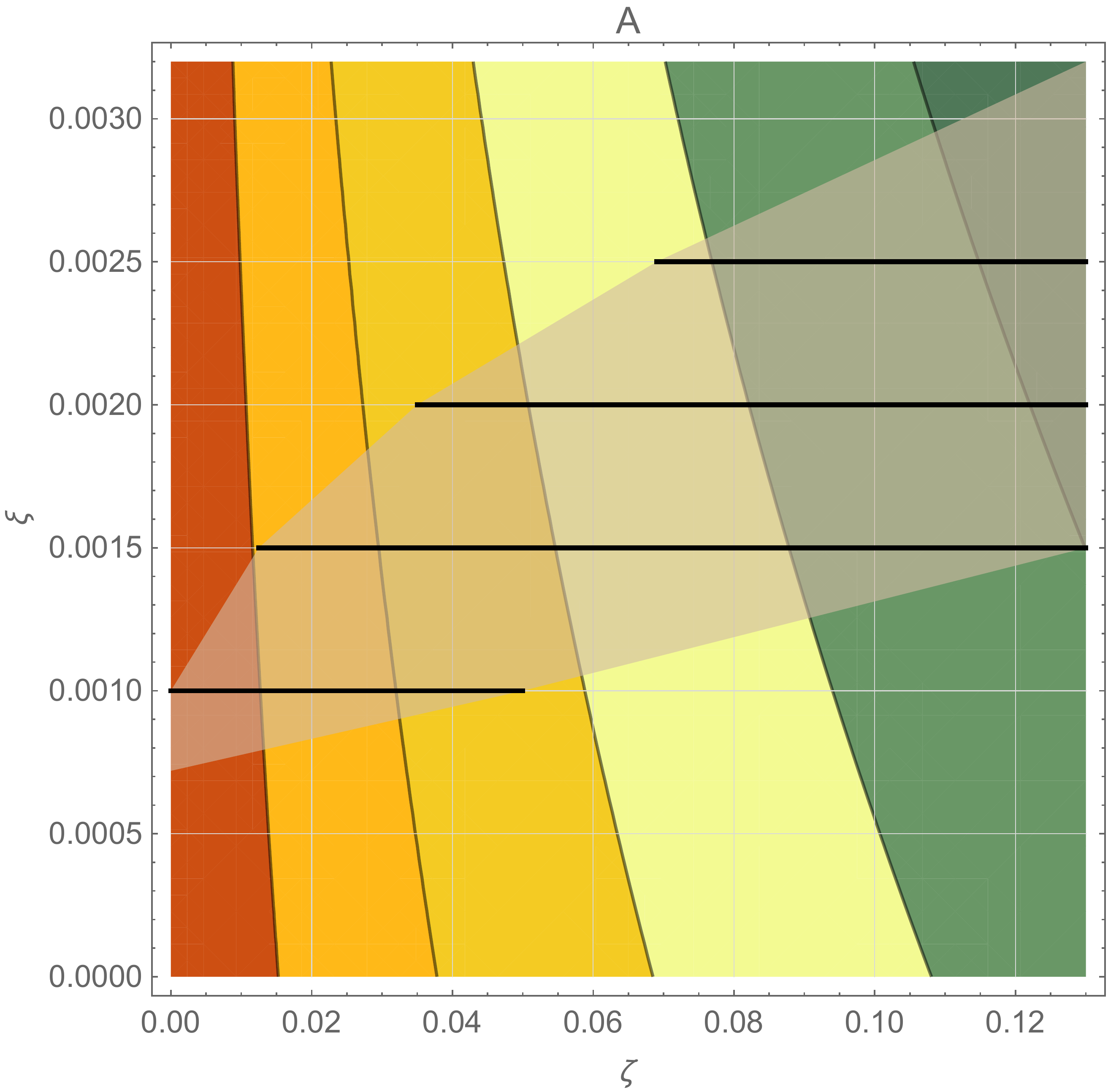}
\includegraphics[width=0.03\textwidth]{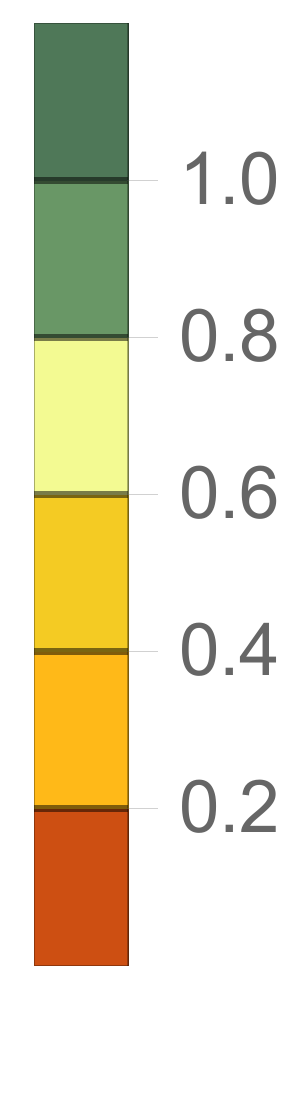}
\end{center}
\begin{center}
\includegraphics[width=0.28\textwidth]{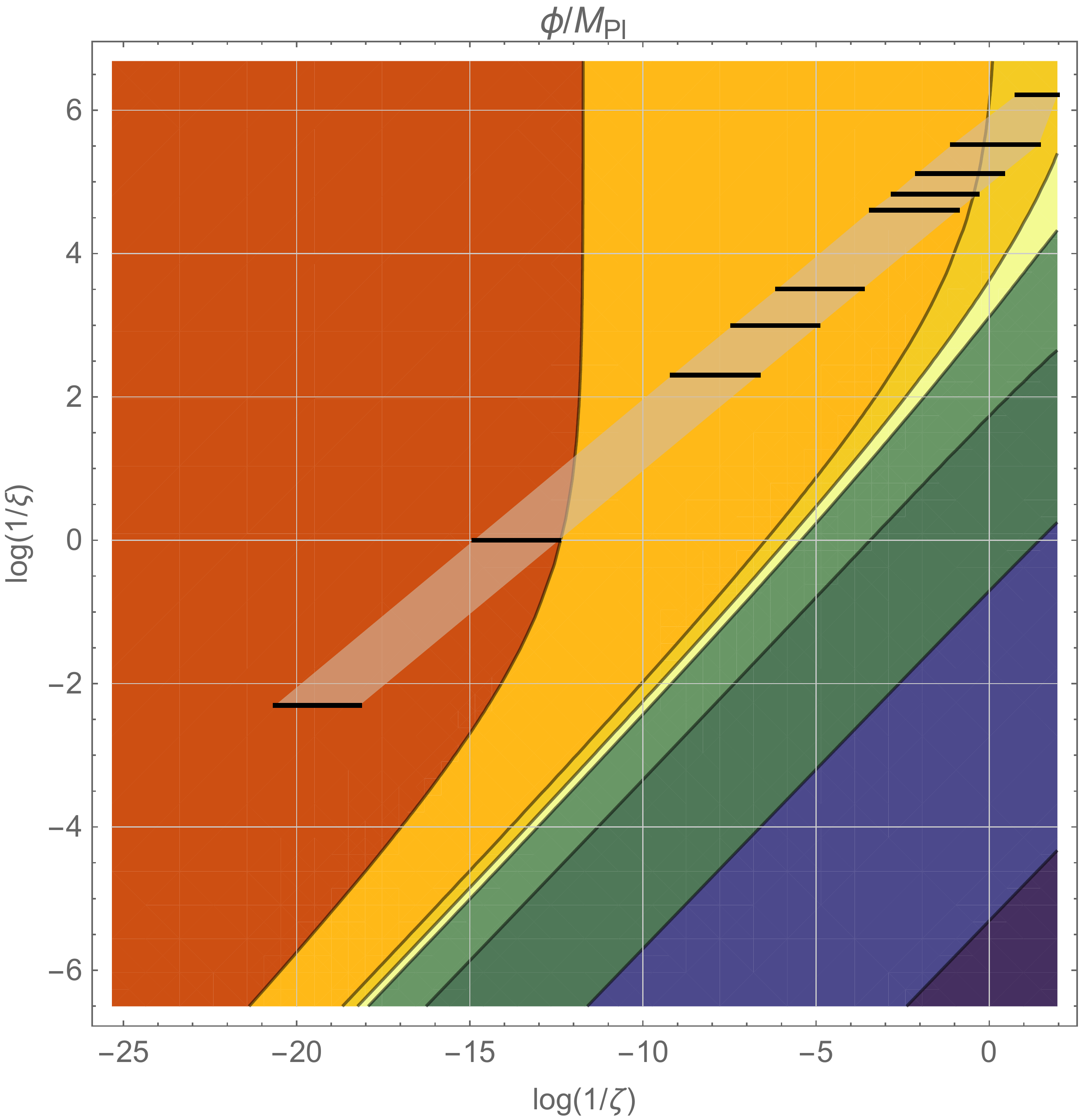}
\includegraphics[width=0.03\textwidth]{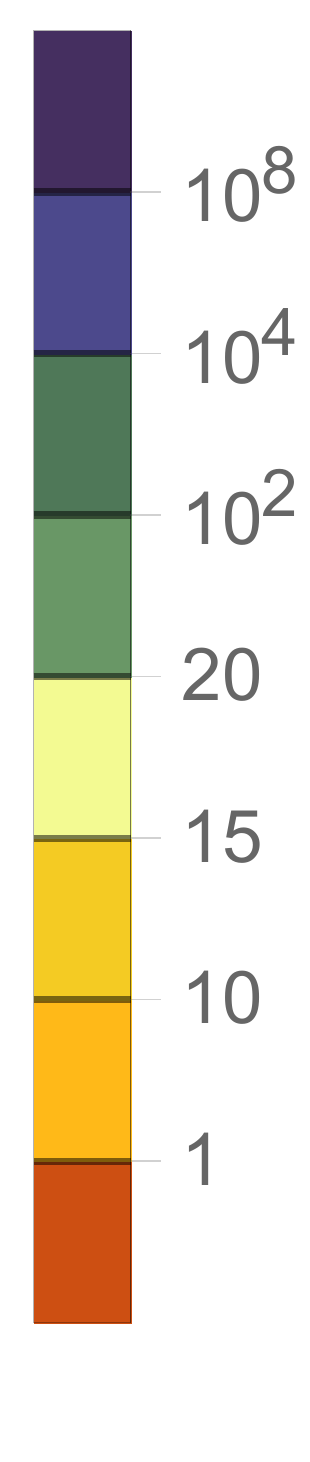}
\includegraphics[width=0.28\textwidth]{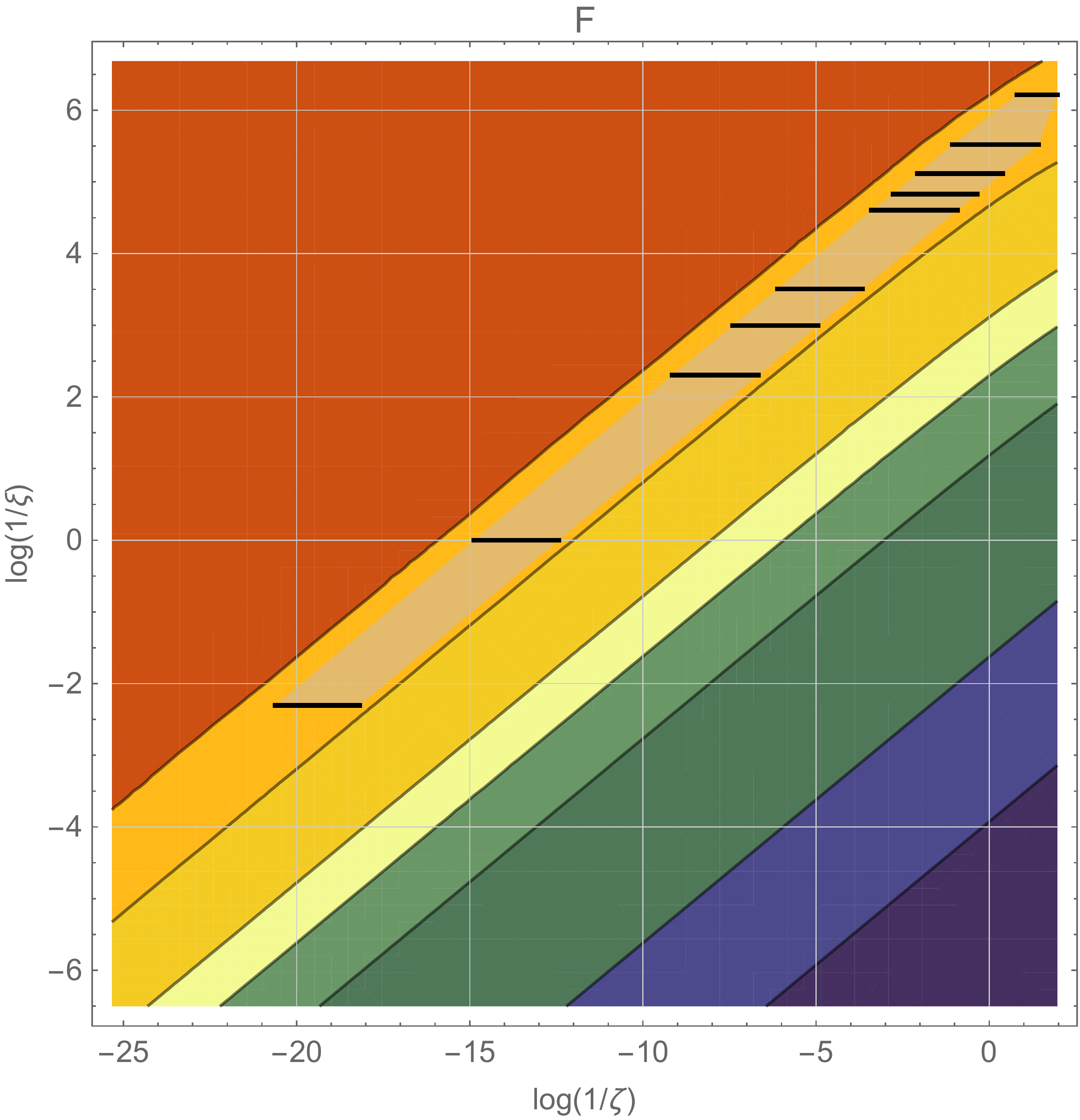}
\includegraphics[width=0.03\textwidth]{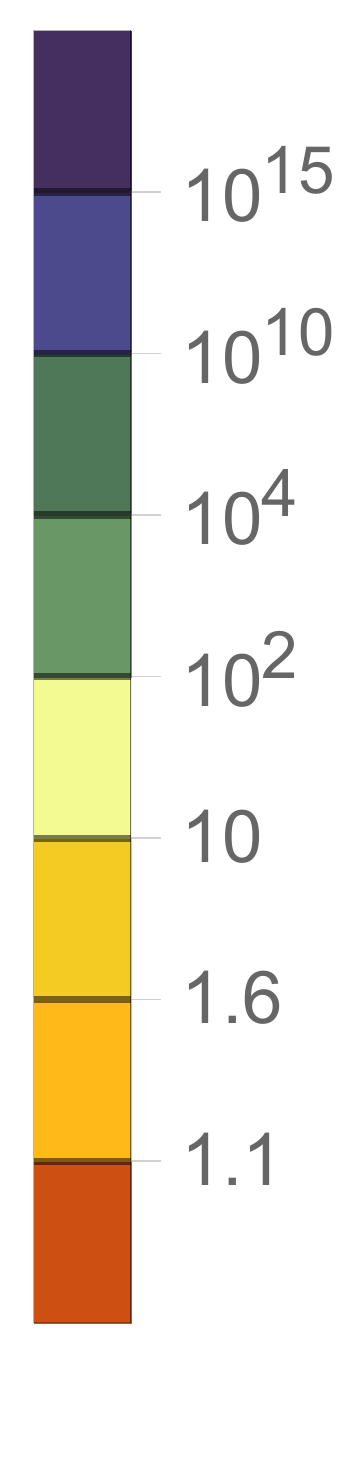}
\includegraphics[width=0.28\textwidth]{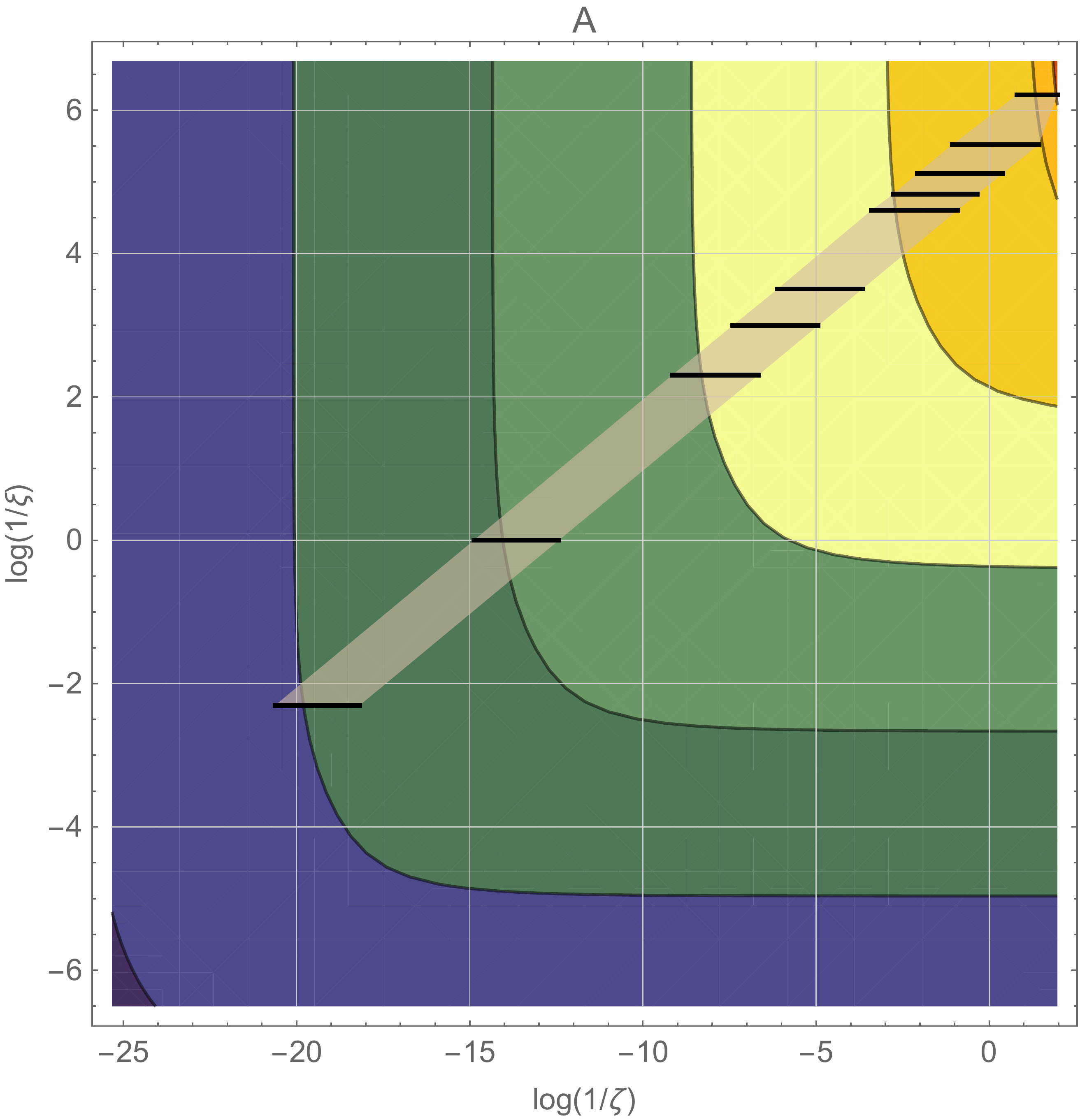}
\includegraphics[width=0.03\textwidth]{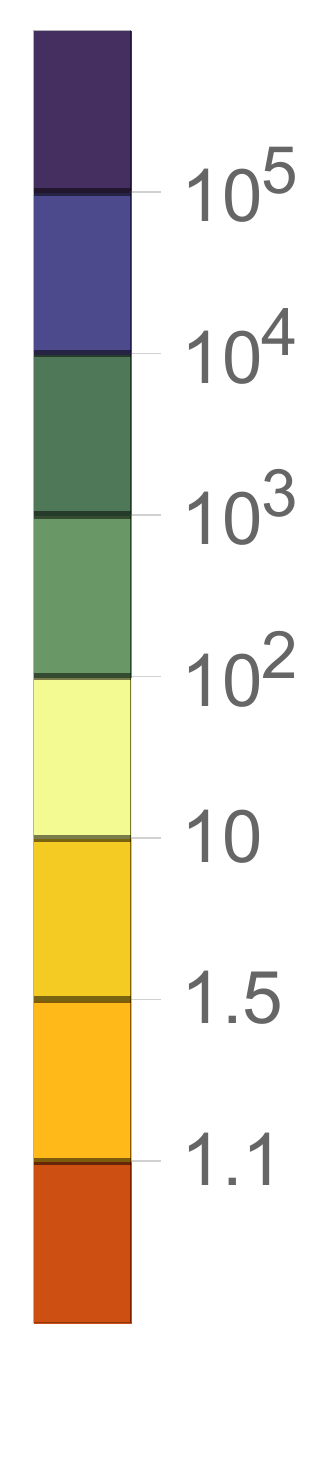}
\caption{\it{We depict the behaviour of $x=\phi/M_{pl}$, $F$ and $\mathcal{A}$ for the solution $x^{(1)}$ with $\zeta<0.13$ (upper row) and for the other one $x^{(2)}$ with $\zeta>0.13$ (lower row) in the space of parameters $\xi$ and $\zeta$, for the power $n=2$ and the number of $e$-folds fixed to $N=70$. The gray region is built using the physical ranges (horizontal black lines) of the parameters $\xi$ and $\zeta$ in Table \ref{Table1} and Table \ref{Table2}, which have been determined by observational bounds in the $n_s-r$ plane (FIG. \ref{r_ns_Galileon}).}}
\label{FIG_Galileon_x}
\end{center}
\end{figure*}

Moreover, by using Eq. \eqref{SlowRoll-Parameters} the slow-roll parameter $\delta_{GX}$ is given by
\bea
\delta_{GX}\equiv \dfrac{-\dot{\phi}X G_{,X}}{M_{pl}^2 H F}=\frac{\mathcal{A}}{6 F}\left(\frac{\dot{\phi}}{M_{pl} H}\right)^2 \simeq \frac{\mathcal{A}}{3}\delta_{PX}.
\label{deltaGX}
\eea Putting the relations \eqref{deltaF2}, \eqref{deltaPX2} and \eqref{deltaGX} into the equation \eqref{epsilon2} for the slow-roll parameter $\epsilon$, one is led to the expression
\be
\epsilon\simeq \left(\epsilon_{V}-\epsilon_{F}\right) \left(\frac{F}{1+\mathcal{A}}\right),
\label{epsilonGF2}
\ee and the end of inflation takes place at $\phi=\phi_{f}$ such that $\epsilon(\phi_{f})= 1$.
In this case, the expression to compute the number of $e$-folds $N$ is found to be
\be
N\equiv \int_{t*}^{t_{f}}{H dt}\simeq \int_{\phi_{f}}^{\phi_{*}}{\frac{1+\mathcal{A}}{\left[\left(\frac{M_{pl} V_{,\phi}}{V}\right)+\left(\frac{M_{pl} F_{,\phi}}{F}\right)\right] F} \frac{d\phi}{M_{pl}}}.
\label{Nefolds2}
\ee 

The parameter $Q_{s}$ in Eq. \eqref{Qs}, and the square of propagation speed of the scalar mode $c_{s}^2$ in Eq. \eqref{csSlow}, are written as
\bea
\label{Qs2}
&& Q_{s}\simeq M_{pl}^2 F\left(1+2\mathcal{A}\right)\delta_{PX}>0, \\
&& c_{s}^2\simeq \frac{\delta_{PX}\left(1+\frac{4}{3}\mathcal{A}\right)+2 \delta_{F}}{\delta_{PX}\left(1+2 \mathcal{A}\right)}.
\label{csSlow2}
\eea  Thus, the condition $c_{s}^2>0$ is translated to $\delta_{F}/\delta_{PX}>-1/2-2\mathcal{A}/3$.

Also, the power spectra of scalar perturbations, Eq. \eqref{Power_Spectrum_Scalar}, is given by
\bea
&& \mathcal{P}_{s}\simeq \left[\frac{V}{ 24 \pi^2 M_{pl}^4 F^{2}}\right]\frac{\left(\delta_{PX}\left(1+2 \mathcal{A}\right)\right)^{1/2}}{\left(\delta_{PX}\left(1+\frac{4}{3}\mathcal{A}\right)+2 \delta_{F}\right)^{3/2}},
\label{PowerSpectrum_with_Galileon}
\eea being that we also have used \eqref{slowGF1}. The corresponding expressions for the scalar spectral index and the tensor-to-scalar ratio
may be obtained by replacing Eqs. \eqref{deltaGX}, \eqref{epsilonGF2}, and \eqref{csSlow2},
in  Eqs. \eqref{ns} and \eqref{r2}, respectively (not shown).

Similarly as before, in the following we are going to develop a example by considering a specific expression for both the non-minimal coupling function $F$ and the inflaton potential $V(\phi).$

\subsection{Chaotic inflation}

We assume a power-law potential \eqref{PowerLawPotential} and the non-minimal coupling function in \eqref{PowerLawNMF}. 
By using Eq. \eqref{A1}, one finds that the Galileon term $\mathcal{A}(x)$ becomes 
\be
\mathcal{A}(x)=\frac{1}{2} \left[-1+\sqrt{1+4 \zeta  x^{n-1} \left(\frac{2 \xi  x^2}{n \xi  x^2+2 n}+1\right)}\right],
\label{Galileon_Term}
\ee where for convenience we have introduced the dimensionless parameter $\zeta\equiv (\gamma/M_{pl})\times \lambda$.
In this case, the slow-roll parameters $\epsilon_{V}$ and $\epsilon_{F}$ are given by \eqref{epsilonV2} and \eqref{Chao_epF}, respectively. Moreover, the slow-roll parameter $\epsilon$ takes a more complicated form due to the presence of the Galileon term, yielding
\bea
&& \epsilon(x)=\frac{n^2 \left(\xi  x^2+2\right)^2-4 \xi ^2 x^4}{4 x^2 (\mathcal{A}+1) \left(\xi  x^2+2\right)}.
\label{epsilon_with_Galileon2}
\eea Thus, inflation ends when $\epsilon(x)=1$ at $x=x_{f}$. Since we have in this case a polynomial equation of degree seven in the variable $x$, we cannot write below explicit analytical solutions for $x_{f}$. However, by solving it numerically, we find two physical solutions for $x_{f}$: the first solution denoted by $x_{f}=x^{(1)}_{f}$ is defined for $\zeta<0.13$, whereas that the second one $x_{f}=x^{(2)}_{f}$ exists for $\zeta>0.13$. The number of $e$-folds $N$ is calculated from \eqref{Nefolds2}, obtaining
\be
N\simeq 2 \int_{x_{f}}^{x_{*}}\frac{x (\mathcal{A}+1)}{(n+2) \xi  x^2+2 n} \, dx,
\ee where $\mathcal{A}(x)$ is given by Eq. \eqref{Galileon_Term}. Here we may integrate numerically the above equation to obtain the solution for $x$ when a given perturbation scale leaves the Hubble-radius, i.e. $x_{*}$, in terms of $N$ and the others parameters. In doing so, we set the initial conditions at 
the time Hubble-radius.

\begin{figure}[htb]
\begin{center}
\includegraphics[width=0.8\textwidth]{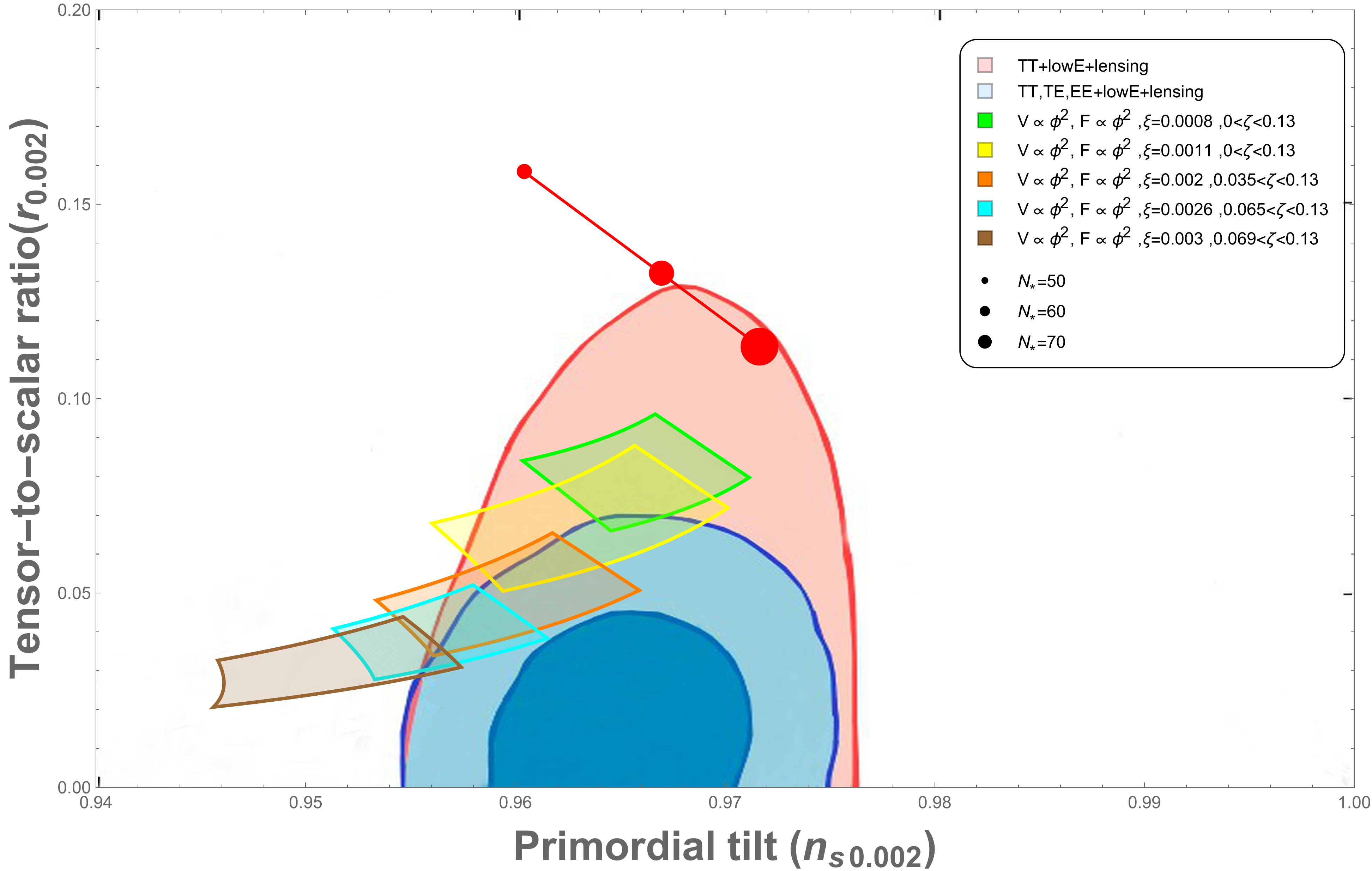}
\includegraphics[width=0.8\textwidth]{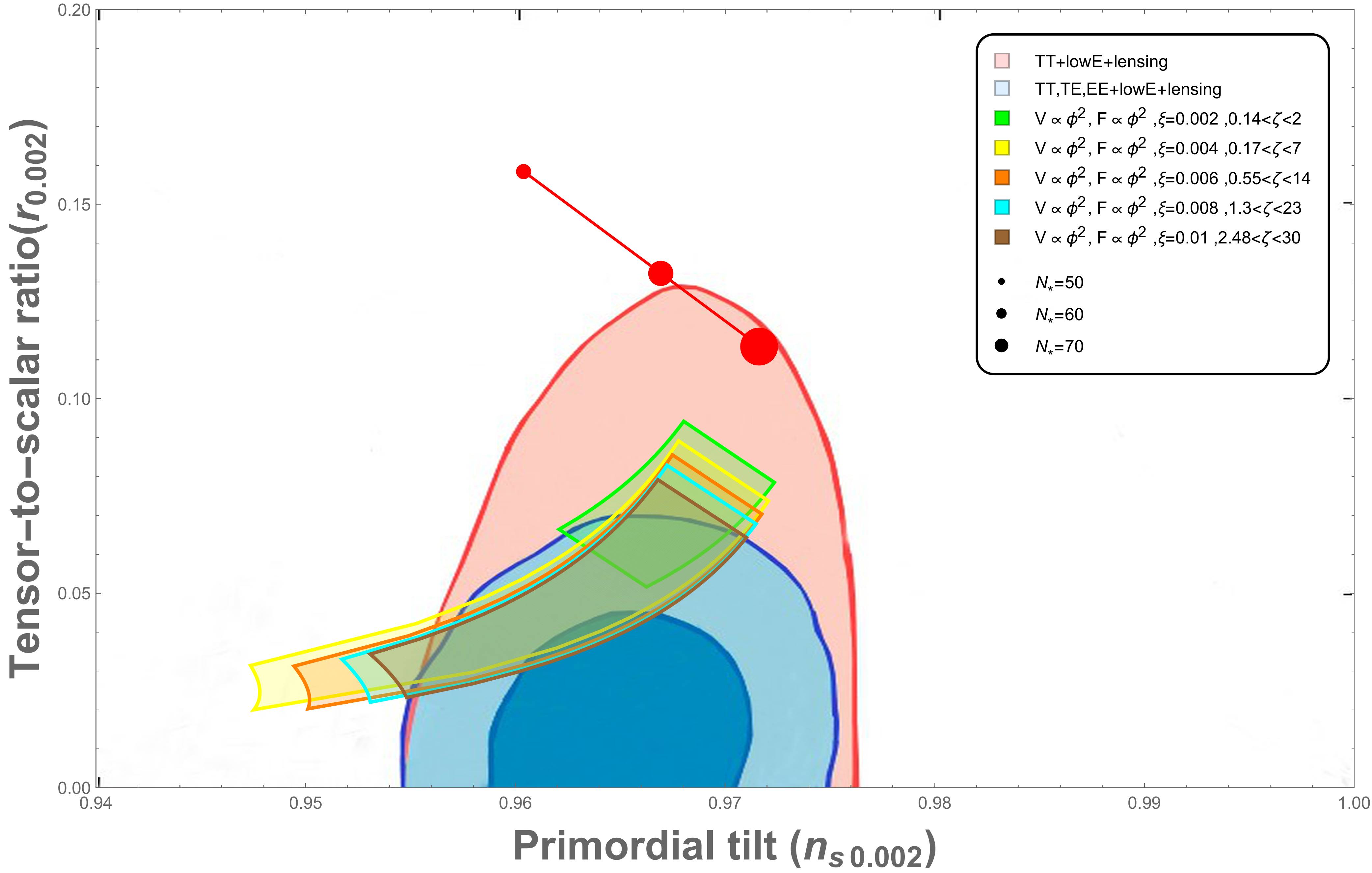}
\caption{\it{We show the plot of the tensor-to-scalar ratio $r$ versus the scalar spectral index $n_s$ in the presence of both, non-minimal coupling to torsion and Galileon interaction term, for quadratic inflation $n=2$ and $N=60-70$. In the upper graph it is shown the solution $x^{(1)}$ whose predictions are superimposed with the $95\%$ C.L region of Planck 2018 data \cite{Akrami:2018odb} and for the which it is obtained the constrained $7.2\times 10^{-4}<\xi< 3.17\times 10^{-3}$ for $N=70$. In the lower graph we shown the solution $x^{(2)}$ which is in good agreement with Planck data at the $68\%$ and $95\%$ C.L. regions. In this case we find the lower bound $\xi> 0.003$ for $N=60$ and $\xi> 0.0014$ for $N=70$.}}
\label{r_ns_Galileon}
\end{center}
\end{figure}

In FIG \ref{FIG_Galileon_x} we plot the numerical solution for $x_{*}$ (left graphic), along with the functions $F$ (centre graphic) and $\mathcal{A}$ (right graphic), in terms of the parameters $\xi$ and $\zeta$, for the particular case of the power $n=2$ and the number of $e$-folds fixed to be $N=70$. The upper row corresponds to the solution $x=x^{(1)}$ which is obtained in the case of solution $x_{f}=x^{(1)}_{f}$, whereas that the lower row is due to the solution $x=x^{(2)}$, associated with $x_{f}=x^{(2)}_{f}$. The gray region indicates the physical region and is determined by using the ranges (horizontal black lines) of the parameters $\xi$ and $\zeta$. These ranges of parameters are found from observational bounds in the $n_s-r$ plane (see FIG. \ref{r_ns_Galileon}) and they are shown in Table \ref{Table1} and \ref{Table2}. One may observe that for the solution $x^{(1)}$, a trans-Planckian
displacement of the inflaton field through the potential, with  $x$ between $x\gtrsim 13 $ and $x\lesssim 18$, as usually happens in large-field models. This solution always works in an intermediate regime of no-nminimal coupling between $F-1=\xi x^2/2 \ll 1$ and $F-1=\xi x^2/2 \gg 1$, with $F$ taking values within the range $1.1< F <1.4$. Furthermore, this solution allows that inflation takes places
in a regime where the Galileon self-interaction becomes sub-dominant in comparison to
the standard kinetic term with $\mathcal{A}\ll 1$ or in an intermediate regime between $\mathcal{A}\ll 1$ and $\mathcal{A}\gg 1$, being that $\mathcal{A}$ assumes values in the range $0<\mathcal{A}<1.2$. In the case of solution $x^{(2)}$, the inflaton field satisfies $0<x\lesssim 15$, and therefore allowing both
trans-Planckian and sub-Planckian values for the inflaton field. So, the no-nminimal coupling to gravity always operates in an intermediate regime between $F-1=\xi x^2/2 \ll 1$ and $F-1=\xi x^2/2 \gg 1$, with $1.1<F<1.6$. Additionally, the Galileon self-interaction works in an intermediate regime between $\mathcal{A}\ll 1$ and $\mathcal{A}\gg 1$ and a strong regime with $\mathcal{A}\gg 1$, such that $1.1 \lesssim \mathcal{A}\lesssim 10^4$. Thus, for this solution sub-Planckian values are reached as long as $\mathcal{A}\gg 1$.
From Eq. \eqref{PhiG}, for $n=2$, one finds that the transition from the regime $\mathcal{A}> 1$ to the regime $\mathcal{A}< 1$ occurs for
\bea
&& x_{G}=\frac{\frac{2 \left(\xi -3 \zeta ^2\right)}{\sqrt[3]{\mathcal{G}}}+\frac{2 \sqrt[3]{\mathcal{G}}}{\xi }+2}{6 \zeta},
\eea
where
\bea
&& \mathcal{G}=\frac{3  \sqrt{3}}{2} \zeta  \xi ^{3/2} \sqrt{4 \zeta ^4+71 \zeta ^2 \xi +8 \xi ^2}+\xi ^3+\frac{45 \zeta ^2 \xi ^2}{2}.
\eea In the case of solution $x^{(1)}$ it is satisfied the condition $x_{f}/x_{G}<1$ for the all the range of values of parameters $\xi$ and $\zeta$, and hence, the end of the regime dominated by Galileon self-interaction always takes place during slow-roll inflation. On the other hand, for solution $x^{(2)}$, we find $x_{f}/x_{G}<1$ for $0.13<\zeta<2$, and $x_{f}/x_{G}>1$ for $\zeta\gtrsim 2$. So, for $0.13<\zeta<2$ the behaviour is similar to the case of solution $x^{(1)}$, whereas that, for $\zeta\gtrsim 2$ the dominance of Galileon self-interaction is extended up to after the end of slow-roll inflation which could spoils the oscillatory regime of inflaton with a negative propagation speed squared $c^2$ of the scalar mode, leading to Laplacian instabilities \cite{Ohashi:2012wf}.

With Eqs. \eqref{deltaF2},\eqref{deltaPX2} and \eqref{dotPhi_Galileon}, we obtain 
\be
\delta_{F}=-\frac{\xi \left(n+\frac{2 \xi  x^2}{2+\xi  x^2}\right)}{1+\mathcal{A}},
\ee and 
\be
\delta_{PX}=\frac{\left[(n+2) \xi  x^2+2 n\right]^2}{4 x^2 (\mathcal{A}+1)^2 \left(\xi  x^2+2\right)}.
\ee

\begin{figure*}[htb]
\begin{center}
\includegraphics[width=0.28\textwidth]{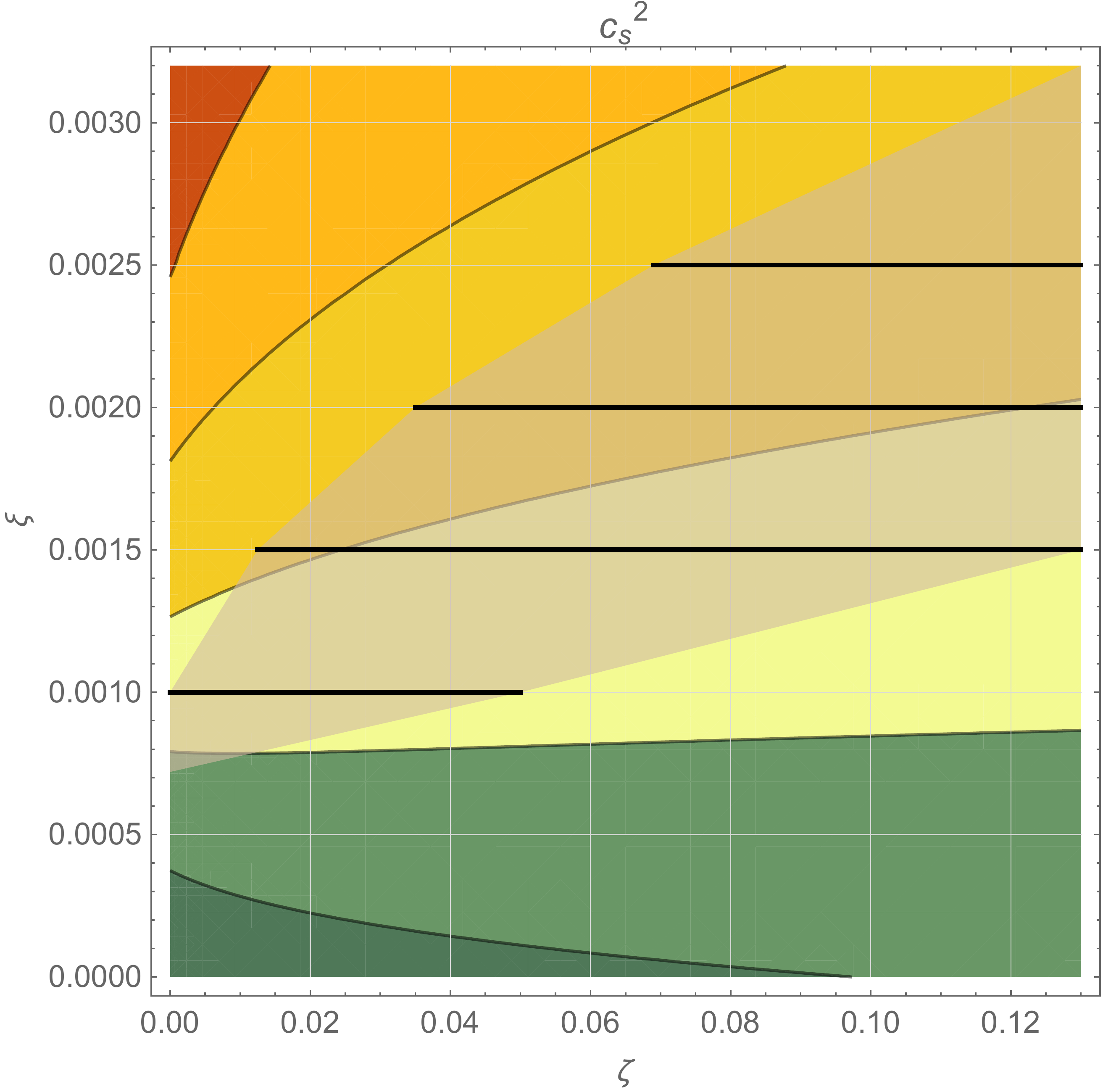}
\includegraphics[width=0.03\textwidth]{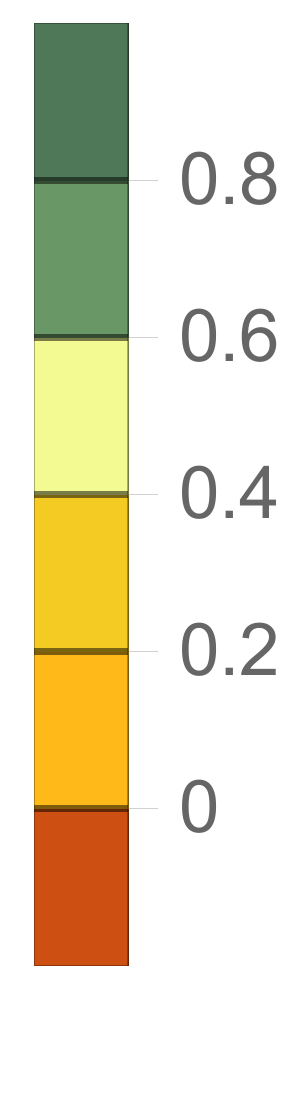}
\includegraphics[width=0.28\textwidth]{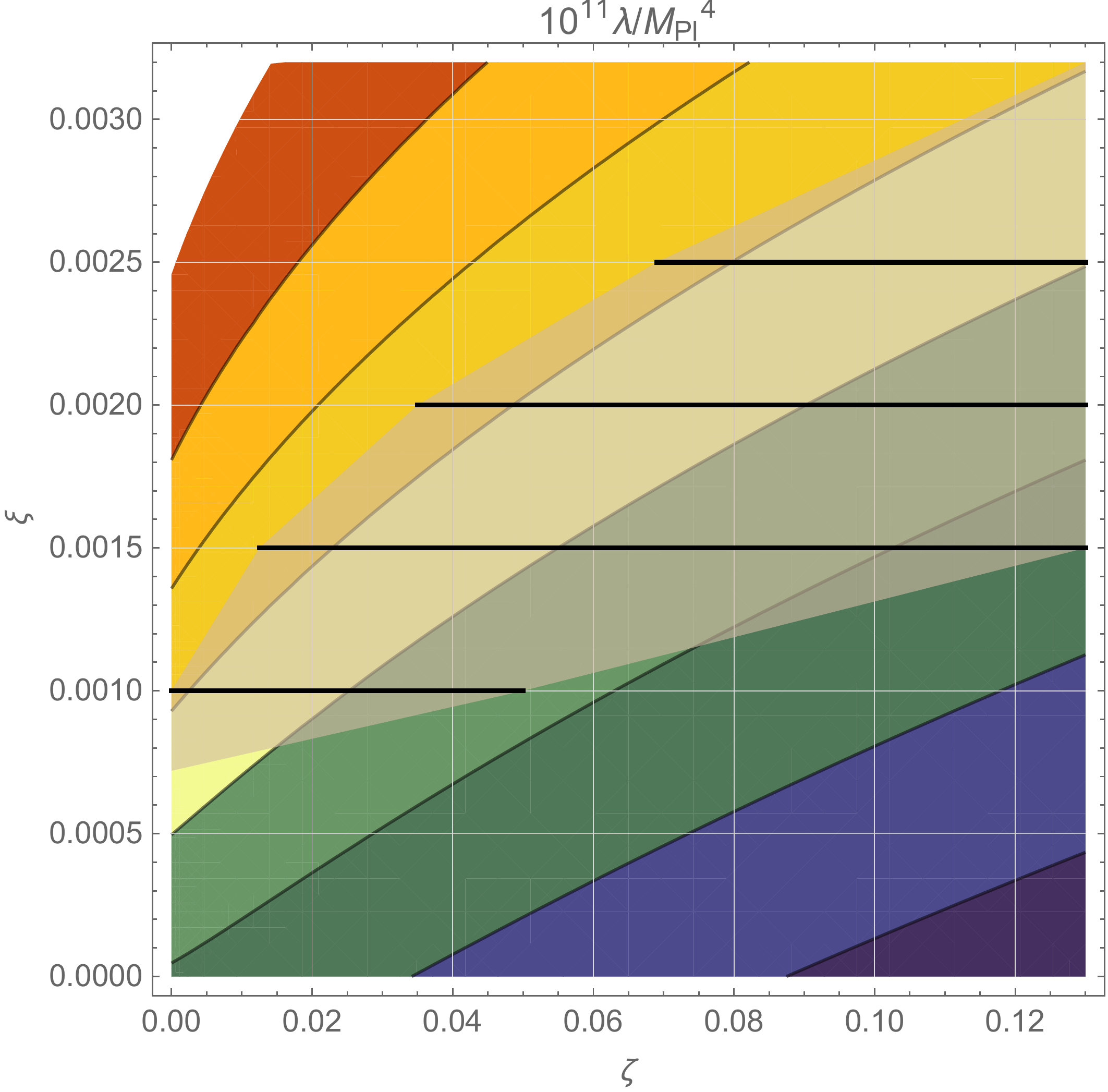}
\includegraphics[width=0.03\textwidth]{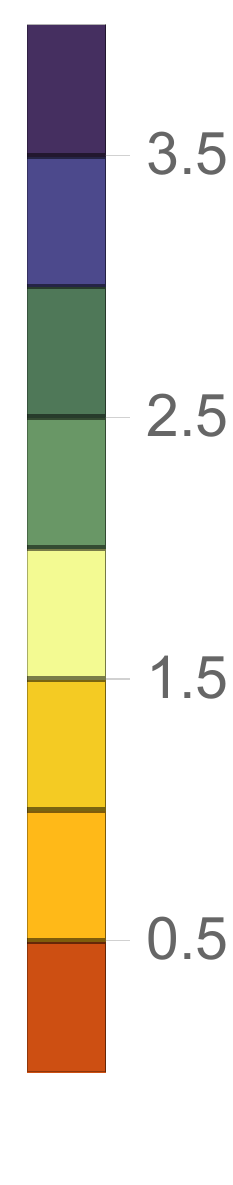}
\includegraphics[width=0.28\textwidth]{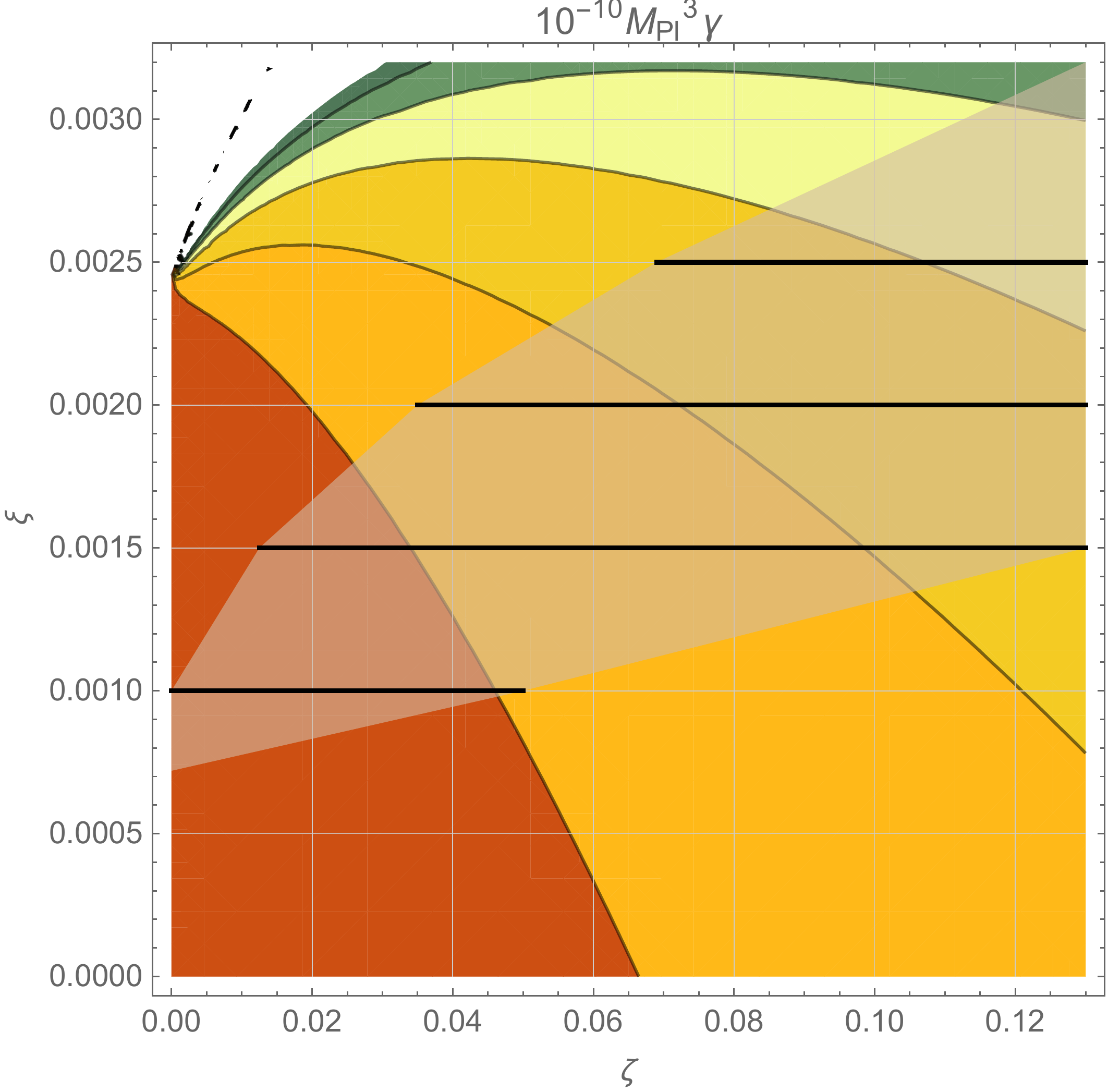}
\includegraphics[width=0.03\textwidth]{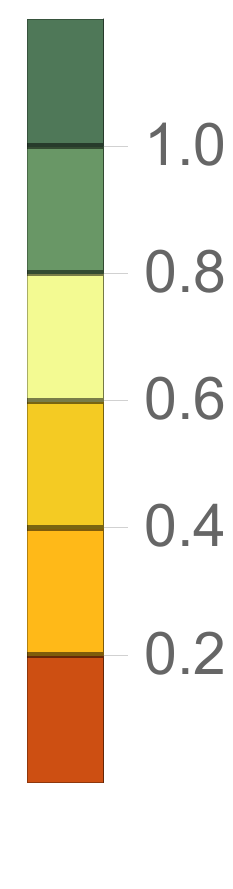}
\end{center}
\begin{center}
\includegraphics[width=0.28\textwidth]{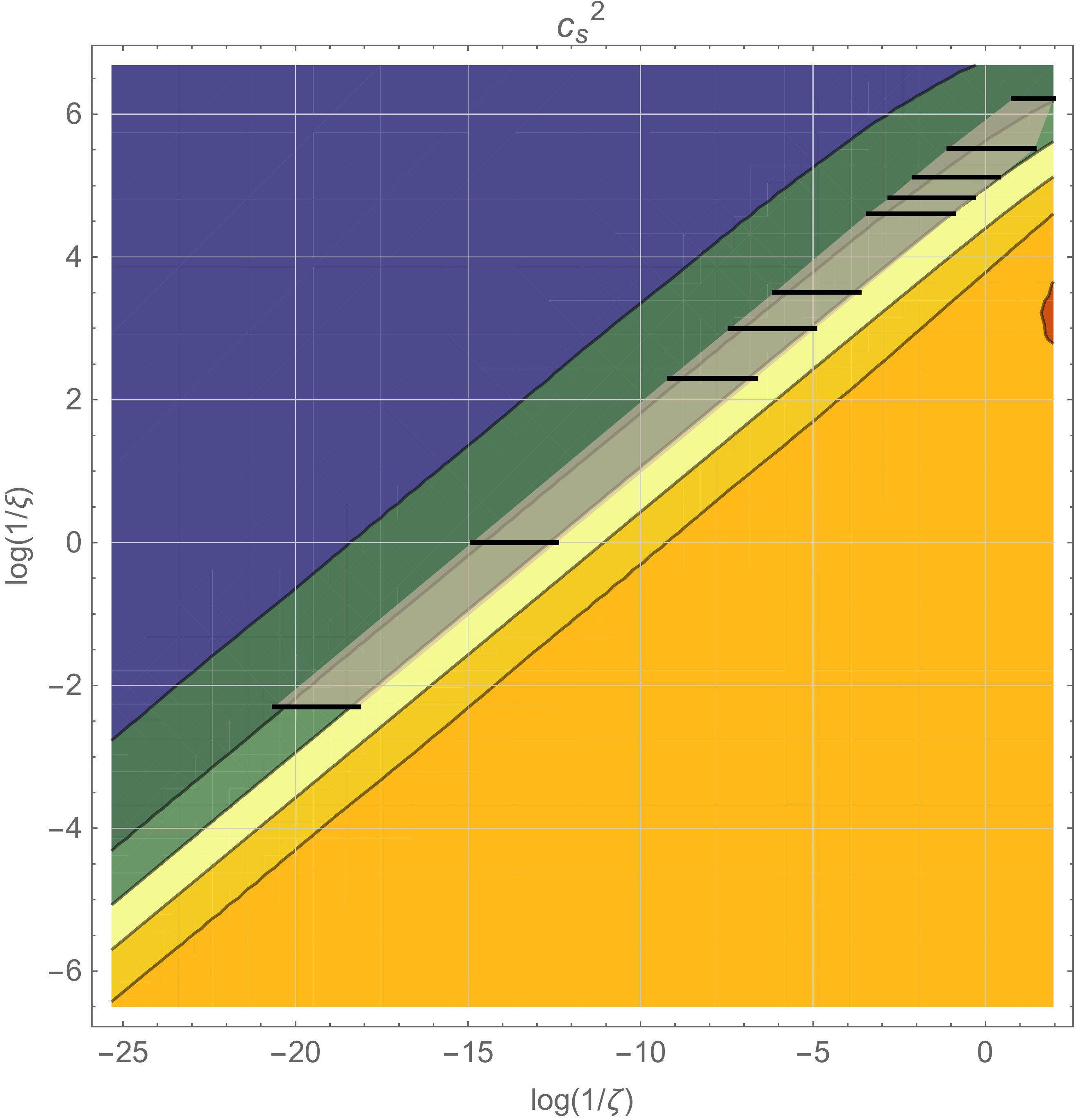}
\includegraphics[width=0.03\textwidth]{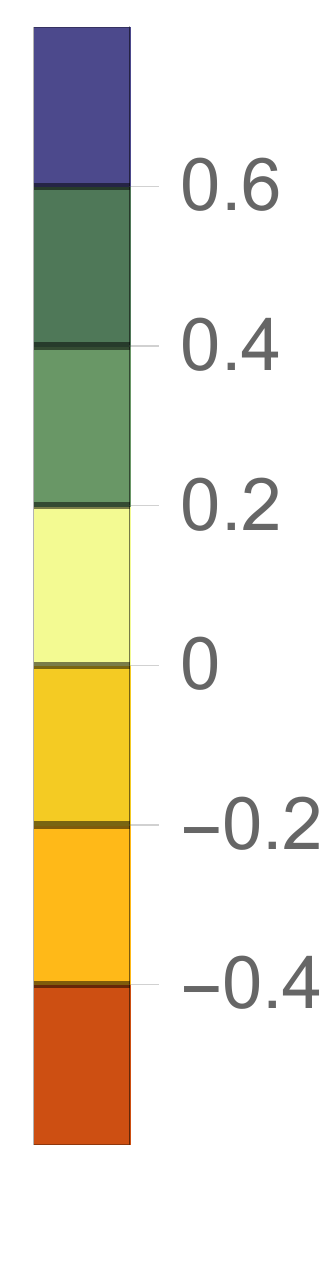}
\includegraphics[width=0.28\textwidth]{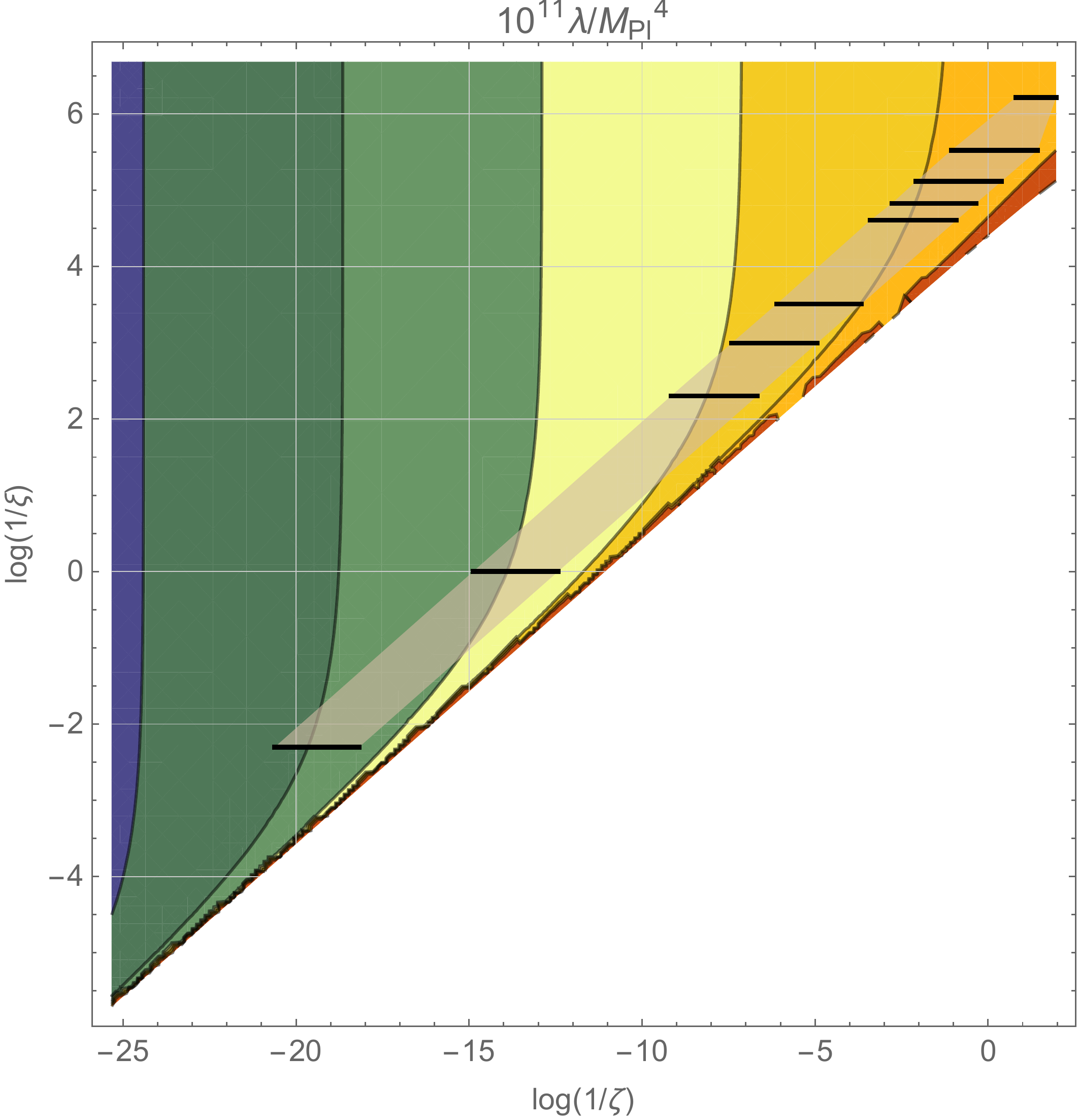}
\includegraphics[width=0.03\textwidth]{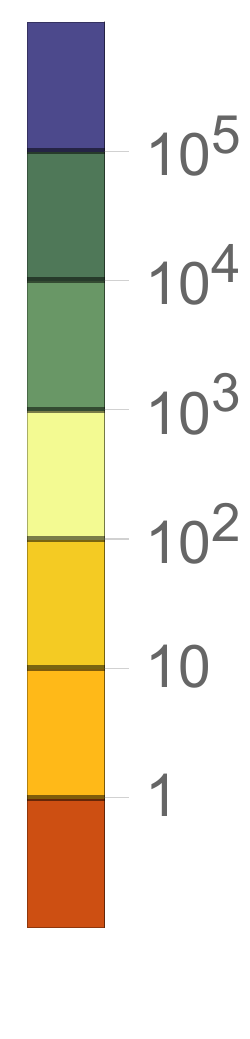}
\includegraphics[width=0.28\textwidth]{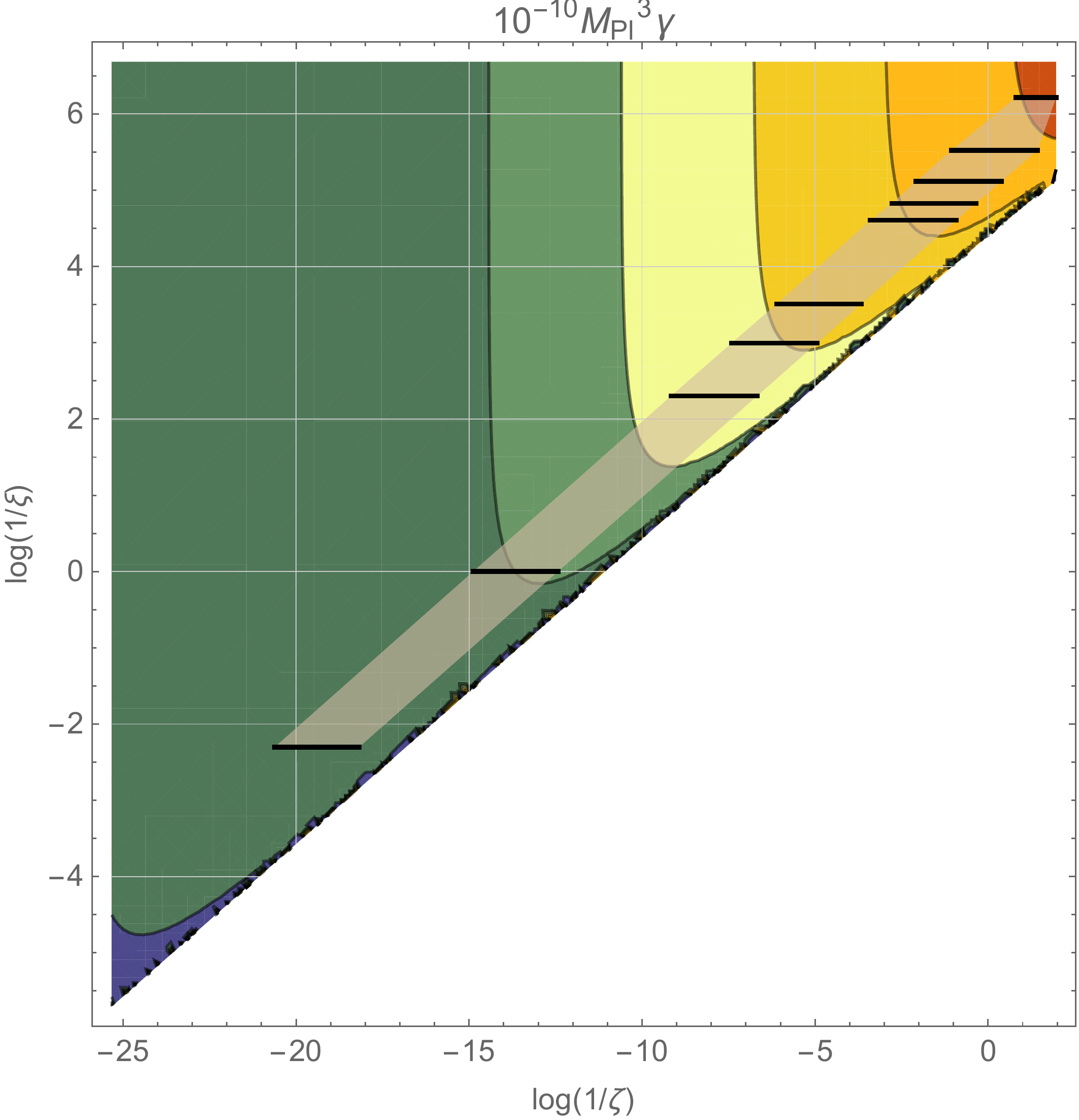}
\includegraphics[width=0.03\textwidth]{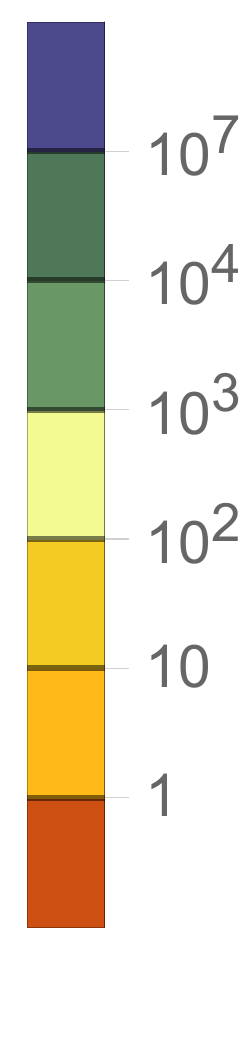}
\caption{\it{We depict the behaviour of scalar propagation speed squared $c_{s}^2$, parameters $\lambda$ and $\gamma$ consistent with scalar power spectrum $P_{s}=2.141\times 10^{-9}$, in the case of solution with $\zeta<0.13$ (upper row) and for the other one with $\zeta>0.13$ (lower row), in the plane of parameters  $\xi$ and $\zeta$, for $n=2$ and $N=70$. The gray region is built using the physical ranges (horizontal black lines) of the parameters $\xi$ and $\zeta$ in Table \ref{Table1} and Table \ref{Table2}, which have been determined by observational bounds in the $n_s-r$ plane (FIG. \ref{r_ns_Galileon}).}}
\label{cs_Galileon}
\end{center}
\end{figure*}

Thus, using the above expressions and Eq. \eqref{deltaGX} into Eq. \eqref{csSlow2}, it is easy to obtain the field dependence of scalar propagation speed squared as
\be
c_{s}^2=\left[\frac{1}{1+2 \mathcal{A}}\right]\left[1+\frac{4 \mathcal{A}}{3}-\frac{8 \xi  x^2 (\mathcal{A}+1)}{(n+2) \xi  x^2+2 n}\right].
\label{cs2_Galileon} 
\ee  Also, the power spectrum $\mathcal{P}_{s}$ in Eq. \eqref{PowerSpectrum_with_Galileon} becomes
\bea
\mathcal{P}_s&=&\frac{2 \sqrt{3} \tilde{\lambda}  \sqrt{1+2 \mathcal{A}}\left(1+\mathcal{A}\right)^2 x^{2+n}}{n \pi ^2 \left(2+\xi  x^2\right) \sqrt{2 n+(2+n) \xi  x^2}}\nonumber\\
&&\times\left[6 n+3 (n-6) \xi  x^2+4 \mathcal{A} \left(2 n+(n-4) \xi  x^2\right)\right]^{-\frac{3}{2}}.\label{PGf}
\eea 
After evaluating \eqref{PGf} at the value of the scalar field when a given perturbation scale leaves the Hubble-radius, which was already found numerically, and by using the current observational value for the amplitude of primordial scalar perturbations $\mathcal{P}_{s}= 2.141\times 10^{-9}$ \cite{Ade:2015lrj,Akrami:2018odb}, we may find a constraint for $\tilde{\lambda}\equiv \lambda/M_{pl}^4$ in terms of $\xi$ and $\zeta$. Thus, by using the values of $\tilde{\lambda}$ and $\zeta$ we recover the values of $\tilde{\gamma}\equiv\gamma M_{pl}^3$. 

In FIG \ref{cs_Galileon} it is shown the behaviour of $c_{s}^2$ (left graphic), $\lambda$ (centre graphic) and $\gamma$ (right graphic), in the $\zeta-\xi$ plane for $n=2$ and $N=70$. As before, the gray region indicates the physical region of the parameters $\xi$ and $\zeta$, as it has been obtained from the observational bounds in the $n_s-r$ plane (FIG \ref{r_ns_Galileon}), whose corresponding ranges are summarized in Tables \ref{Table1} and \ref{Table2}. For both solutions $x^{(1)}$ (upper row) and $x^{(2)}$ (lower row) one finds that $c_{s}^2$ is always positive and therefore there is no Laplacian instabilities
during inflation. For the solution $x^{(1)}$ the parameters $\lambda$ and $\gamma$ are constrained to be $1\times 10^{-11}\lesssim \tilde{\lambda}\lesssim 3.0\times 10^{-11}$ and $\tilde{\gamma} \lesssim  8 \times 10^{9}$. In the case of $x=x^{(2)}$ (lower row), one is free of this type of pathology above the straight line $\log(1/\xi)=4.34 + 0.397\log(1/\zeta)$, and the parameters $\lambda$ and $\gamma$ are bounded from below by $\tilde{\lambda}\gtrsim 10^{-11}$ and $\tilde{\gamma}\gtrsim 10^{9}$. Therefore, we must constraint from above the parameters $\lambda$ and $\gamma$ with the help of some additional phenomenological considerations, because these quantities may reach arbitrarily large values for this solution and still provide results compatible with the observations.

Substituting Eqs. \eqref{epsilon_with_Galileon2}, \eqref{deltaF_Coupl} and \eqref{deltaGX} into Eq. \eqref{epsilons2}, one finds that the slow-roll parameter $\epsilon_{s}$ is given by
\bea
\epsilon_{s}&\simeq & \epsilon+\delta_{F}+\delta_{GX},\nonumber\\
&=&\frac{n^2 \left(4 \mathcal{A}+3\right)}{6 x^2 (\mathcal{A}+1)^2}+\frac{2\xi (8 \mathcal{A}+9 )}{3 (\mathcal{A}+1)^2 \left(\xi  x^2+2\right)}+\frac{(n+2) \xi \left[4 \left(n-4\right) \mathcal{A}+3\left( n-6\right)\right]}{12 (\mathcal{A}+1)^2}.
\label{epss_Galileon}
\eea Hence, it is straightforward see that 
\bea
\eta_{s}&=&\left[\frac{1}{x^2 (\mathcal{A}+1)^2 \left(\xi  x^2+2\right) }\right]\Bigg[\frac{2 \xi ^2 x^4 (\mathcal{A}+1) \left(4 \left(n^2+8\right) \mathcal{A}+3 \left(n^2+12\right)\right)}{\xi  x^2 (4 (n-4) \mathcal{A}+3 (n-6))+2 n (4 \mathcal{A}+3)}\nonumber\\
&& + \frac{8 n^2 (\mathcal{A}+1)\left(\xi  x^2 +1 \right) (4 \mathcal{A}+3)}{\xi  x^2 (4 (n-4) \mathcal{A}+3 (n-6))+2 n (4 \mathcal{A}+3)} \nonumber\\
&& +\frac{\mathcal{B}}{\xi  x^2 (4 (n-4) \mathcal{A}+3 (n-6))+2 n (4 \mathcal{A}+3)}\Bigg],
\eea where we have defined 
\bea
\mathcal{B}&=&\left[\frac{\zeta x^{n-1}}{n \left(\xi  x^2+2\right) (1+2\mathcal{A})}\right]  \Big[\xi  x^2 (2 (n-4) \mathcal{A}+n-10)+2 n (2 \mathcal{A}+1)\Big]\nonumber\\
&& \times\left[(n+2) \xi  x^2+2 n\right]\Big[\left(n^2+n-2\right) \xi ^2 x^4+ 4 \left(n^2+1\right) \xi  x^2+4 (n-1) n\Big].
\eea
On the other hand, from Eqs. \eqref{cs2_Galileon} and \eqref{etas}, we find
\bea
&& s=\frac{24 n \xi+\frac{\mathcal{C}}{2 n x^2 (\mathcal{A}+1) (2 \mathcal{A}+1)^2 \left(\xi  x^2+2\right)}}{\xi  x^2 (4 (n-4) \mathcal{A}+3 (n-6))+2 n (4 \mathcal{A}+3)},
\eea where
\bea
\mathcal{C}=\frac{n (2 \mathcal{A}+1) \left((n-10) \xi  x^2+2 n\right)\mathcal{B}}{\xi  x^2 (2 (n-4) \mathcal{A}+n-10)+2 n (2 \mathcal{A}+1)}.
\eea
These results lead us to write down the spectral index of $\mathcal{R}$ \eqref{ns} as
\bea
 n_{s}&=& 1+\frac{\xi  \left((n+2) \xi  x^2+2 n\right)}{(\mathcal{A}+1) \left(\xi  x^2+2\right)}-\frac{n^2 \left(\xi  x^2+2\right)^2-4 \xi ^2 x^4}{2 x^2 (\mathcal{A}+1) \left(\xi  x^2+2\right)}\nonumber\\
&& -\frac{24 n \xi+\frac{\mathcal{C}}{2 n x^2 (\mathcal{A}+1) (2 \mathcal{A}+1)^2 \left(\xi  x^2+2\right)}}{\xi  x^2 (4 (n-4) \mathcal{A}+3 (n-6))+2 n (4 \mathcal{A}+3)}\nonumber\\
&&-\frac{\mathcal{D}}{\xi  x^2 (4 (n-4) \mathcal{A}+3 (n-6))+2 n (4 \mathcal{A}+3)}\label{nsAf},
\eea where 
\bea
\mathcal{D}&=&\frac{8 n^2 \left(\xi  x^2 + 1\right) (4 \mathcal{A}+3)}{x^2 (\mathcal{A}+1) \left(\xi  x^2+2\right)} +\frac{2 \xi ^2 x^2 \left[4 \left(n^2+8\right) \mathcal{A}+3 \left(n^2+12\right)\right]}{(\mathcal{A}+1) \left(\xi  x^2+2\right)}\nonumber\\
&&  + \frac{\mathcal{C} \left[\frac{\xi (2 (n-4) \mathcal{A}+n-10)}{n (2 \mathcal{A}+1)}+\frac{2}{ x^2}\right]}{(\mathcal{A}+1)^2 \left(\xi  x^2+2\right) \left((n-10) \xi  x^2+2 n\right)}.
\eea

Finally, from Eqs.  \eqref{r1}, \eqref{cs2_Galileon} and \eqref{epss_Galileon}, we obtain for the tensor-to-scalar ratio $r$ the following expression
\bea
 r&=& \frac{4 (2 \mathcal{A}+1) \left((n+2) \xi  x^2+2 n\right)^2}{3 \sqrt{3} x^2 (\mathcal{A}+1)^2 \left(\xi  x^2+2\right)}\nonumber\\
&&\times\left[\frac{\xi  x^2 (4 (n-4)\mathcal{A}+3 (n-6))+2 n (4 \mathcal{A}+3)}{(2 \mathcal{A}+1) \left((n+2) \xi  x^2+2 n\right)}\right]^{3/2}\label{rAf}.
\eea In a similar fashion as before, after evaluating these inflationary observables at the value of the scalar field when a given perturbation scale leaves the Hubble-radius, we are able to compare the theoretical predictions of this second subclass of model in the $n_s-r$ plane with the allowed contour regions from Planck 2018 data. Then, we find the allowed ranges of the parameters characterizing this subclass of model.

In FIG \ref{r_ns_Galileon} we show the trajectories in the $n_s-r$ plane for our second model,
which are generated by plotting Eqs. \eqref{nsAf} and \eqref{rAf} (after being evaluated at $x_{*}$) parametrically, varying both the number of $e$-folds $N$ and the parameter $\zeta$ in a wide range, for several fixed values of $\xi$ and for the power $n$ fixed to $n=2$.
In addition, we have considered the two-dimensional marginalized joint confidence contours for $(n_s,r)$, at
the $68\%$ and $95\%$ C.L., from the latest Planck data \cite{Ade:2015lrj,Akrami:2018odb}.
In the upper graph it is shown the solution $x^{(1)}$ for which it is found that the predictions of the model are within the $95\%$ C.L. region from Planck 2018 data \cite{Akrami:2018odb} provided
that the non-minimal coupling parameter $\xi$ takes values within the allowed range $7.20\times 10^{-4}\lesssim \xi\lesssim 3.17 \times 10^{-3}$, for $N=70$. On the other hand, $\zeta$ takes values in some specific range which depends on the value of $\xi$. For these values of $\zeta$ we obtain $\tilde{\lambda}\sim 10^{-11}$, $m_{\phi}/ M_{pl}\sim 10^{-6}$, and $\tilde{\gamma}\lesssim 10^{9}$, as it is shown in Table \ref{Table1}. Therefore, the predicted range for the
tensor-to-scalar ratio $r$, for $N=70$, is found to be $0.028\lesssim r\lesssim 0.069 $. Hence,
this suppression on the tensor-to-tensor to scalar ratio, due the presence of both
the non-minimal coupling and the Galileon self-interaction, allows to bring chaotic quadratic to be compatible with current observations, when the first solution $x^{(1)}$ is considered. Now, by considering the expected upper values on $r$ for the forthcoming B-mode polarization experiments, BICEP3 \cite{Wu:2016hul} or LiteBIRD \cite{Suzuki:2018cuy}, we might confirm chaotic inflation in scalar-torsion gravity with both non-minimal coupling and a Galileon self-interaction term or we might rule it out.

Now, in the lower graph we show the predictions in the $n_s-r$ plane for the second
solution $x^{(2)}$, which also provides results in consistency with current observational bounds in a wide range of parameter values. In particular, for $N=60$, a lower limit for $\xi$ is found to be $\xi\gtrsim 0.003$ for $N=60$, while for $N=70$, its lower limit yields $\xi\gtrsim 1.4\times 10^{-3}$. However, the observational data, through the $n_s-r$ plane, does not impose any 
upper bound on $\xi$. For values of $\xi$ such that $\xi> 0.003$, one may always find a range for parameter $\zeta$ which is superimposed over the $68\%$ and $95\%$ C.L. regions, and then leading us to arbitrarily large values of parameters $\lambda$ and $\gamma$. In Table \ref{Table2} we show some physical ranges for the parameters of the model consistent with observations, in the cases
where the number of $e$-folds is fixed to be $N=60$ and $N=70$.

Let us discuss about of this latter behaviour of solution $x^{(2)}$. Regarding slow-roll inflation as a low-energy effective theory, the maximal cut-off $\Lambda$ is fixed by Planck scale, that is to say $\Lambda\lesssim M_{pl}$. Now, in order for this effective theory to remain valid during the Hubble-radius crossing, the minimal cut-off is determined by the inflationary Hubble scale, $\Lambda\gtrsim H$, with the masses of the fields satisfying  $m\lesssim H$ \cite{Baumann:2014nda}. Thus, for $\Lambda\simeq \gamma^{-1/3}$, and $H/M_{pl} \sim 10^{-5} $, it is easy to obtain the condition $\tilde{\gamma}\lesssim 10^{15}$, and so also, $m_{\phi}/ M_{pl}\lesssim 10^{-5}$, or equivalently, $\tilde{\lambda}\lesssim 10^{-10}$. From results shown in Table \ref{Table2}, one may see that the constraint imposed on $m_{\phi}$ is stronger than the imposed on $\gamma$, when determining the upper limit for the non-minimal coupling parameter. In this way, by using the constraint on the mass scale, this allow us to fix the upper bound for $\xi$ to be $\xi \lesssim 1$, with the best predictions for $r$ in the ranges $0.038\lesssim r\lesssim 0.070$ for $N=60$, and $0.022\lesssim r\lesssim 0.063$ for $N=70$. Furthermore, whether we have in account that the condition $\zeta\lesssim 2$ is required in order to avoid that the regime of dominance of Galileon self-interaction goes until after the end of slow-roll inflation, and hence spoiling reheating after inflation \cite{Ohashi:2012wf}, the non-minimal coupling parameter $\xi $ is more severely constrained from above to be $\xi\lesssim 0.008$ for $N=60$ and $\xi\lesssim 0.009$ for $N=70$. Thus, one finds that $10^{-11} \lesssim \tilde{\lambda}\lesssim 10^{-10}$ and $10^{9} \lesssim \tilde{\gamma}\lesssim 10^{10}$, with $r\sim 0.042$ for $N=60$ and $r\sim 0.024$ for $N=70$. Finally, when solution $x^{(2)}$ is taken into account, and depending on the predictions for the tensor-to-scalar ratio for $N=60$ or $N=70$, chaotic inflation in this particular scenario may supported or disproved by near future experiments. In this way, these future experiments allow to test the major single-field slow-roll inflation models.

\section{Concluding Remarks}\label{Concluding_Remarks}

In the context of teleparallel gravity, we investigated a generalized scalar-torsion theory and its implications for inflationary cosmology. In this theory we consider a canonical scalar field $x\equiv \phi/M_{pl}$ non-minimally coupled to scalar torsion $T$, along with a Galileon-type field self-interaction as motivated from Dvali-Gabadadze-Porrati braneworld (DPG) model \cite{Dvali:2000hr} and Horndeski theory \cite{Horndeski:1974wa}. We studied a flat Friedmann-Robertson-Walker (FRW) background for which we have calculated the modified field equations, and then we computed the second order action of scalar and tensor perturbations to compute the power spectra of these primordial fluctuations.

The Lorentz invariance is considered as one of the most fundamental symmetries in physics, on which rests the principles of Einstein's general relativity and the standard model of particle physics \cite{Mattingly:2005re}. Nevertheless, at a sufficiently high-energy scale (Planck scale), it is expected that these two field theories converge into a single unified and quantum-consistent description of nature, under a possible breaking of Lorentz symmetry \cite{Gasperini:1985aw,Colladay:1998fq,Kostelecky:1989jw,Colladay:1996iz}. When speaking about breaking of Lorentz symmetry it is important to distinguish between particle Lorentz transformations and observer Lorentz transformations. The first class of transformations correspond to rotations or boosts of particles or localized field distributions in a given local Lorentz frame, whereas that, the second one are rotations or boosts changing the local Lorentz frame \cite{Kostelecky:2003fs,Bluhm:2014oua}. Thus, although a theory may present local particle Lorentz violation, either spontaneously or explicitly due to the presence of background fields, in order to have a physically meaningful theory, it must maintain the local observer Lorentz covariance \cite{Bluhm:2016dzm,Bluhm:2017pje}. Furthermore, in the case of a gravity theory with explicit Local Lorentz violation, it is well known that when guaranteeing independence with the observer, it may be necessary to restrict the geometry, which also imposes severe restrictions on the dynamics of solutions, and therefore, leading us to undesirable results \cite{Kostelecky:2003fs,Bluhm:2016dzm,Bluhm:2017pje,Bluhm:2014oua}. In the framework of modified teleparallel gravity, the scalar-torsion theories provide an excellent opportunity for studying the impact of local Lorentz violation on the dynamics of inflation in the very early universe. TG is invariant under local Lorentz transformations and diffeomorphisms \cite{Aldrovandi-Pereira-book,Arcos:2005ec}. However, once that we introduce a non-minimal coupling term between the scalar field and torsion, in the form $F(\phi) T$, the coupling function $F(\phi)$ can be identified as a background field which introduces an explicit local Lorentz violation into the theory. But, here let us to note that in the FRW background the homogeneity allows us to restore the observer local Lorentz invariance, without having to restrict the geometry. More interestingly, as we have highlighted in section \ref{Coupling}, in the FRW background, the non-minimally coupled scalar-torsion theory (without Galileon) becomes equivalent to the scalar-vector-tensor theory in Ref. \cite{Kanno:2006ty}, where the particle local Lorentz symmetry is spontaneously broken due to a time-like vector field. Thus, we also have shown in section \ref{GS_torsion_gravity}, the particle Lorentz violating stage with large contribution to inflation occurs at scales well deep inside the horizon, being that the modifications to the inflationary observables at the horizon crossing would be product of the reminiscent effect of this (particle) Lorentz violation stage \cite{Lim:2004js}. On the other hand, as it also has been shown in section \ref{GS_torsion_gravity}, at the level of first order cosmological perturbations for small deviations from homogeneity, the violation of observer local Lorentz invariance is no longer hidden by the symmetries of the background, but, it may be restored by the dynamics of the background and the scalar perturbations, in the limit of subhorizon and superhorizon scales. Therefore, this allows us to perform the standard quantization procedure on the second order action for scalar perturbations, for obtaining the scalar power spectrum \cite{DeFelice:2011zh,DeFelice:2011uc}. For tensor perturbations the theory is observer local Lorentz invariant and there is no contribution from additional degrees of freedom (see also section \ref{GS_torsion_gravity}).

In order to obtain concrete results we have firstly studied only the effects of the non-minimal coupling on the dynamics of inflation. We assumed the chaotic potential $V(x)=\lambda x^n/n$ with a non-minimal coupling term which has the form $F(x) T$ with $F(x)=1+\xi x^2/2$. In the slow-roll approximation and by using the latest Planck data \cite{Akrami:2018odb}, we have constrained the coupling parameter $\xi$ to be $10^{-4}\lesssim \xi\lesssim 10^{-3}$ with $10^{-11}\lesssim \lambda/M_{pl}^4\lesssim 10^{-10}$ for $2/3 \leq n\leq 2$. For the particular case of chaotic quadratic inflation $n=2$ we found that the prediction for the tensor-to-scalar ratio is $0.05\lesssim r\lesssim 0.07$, for $7.1\times 10^{-4}\lesssim \xi\lesssim 1.1 \times 10^{-3}$, with $e$-folds number $N=70$, which puts it only within the
$95\%$ C.L. contour. In Ref. \cite{Tenkanen:2017jih} it has been studied quadratic inflation with non-minimal curvature-matter coupling, where the constraint on the tensor-to-scalar ratio
was found to be $0.01\lesssim r<0.12$, with the corresponding non-minimal coupling parameter satisfying $1\times 10^{-3}\lesssim \xi_{R}\lesssim 5\times 10^{-3}$, and $N=60$. Thus, by comparing with our results one may see that in the case of non-minimal coupling to torsion it is required a weaker non-minimal coupling to gravity in order to satisfy the current observational constraints, but in contrast it is necessary a greater amount of inflation than in the case with non-minimal coupling to curvature. However, in both cases, either non-minimal coupling to curvature or torsion, it is possible to rescue the quadratic inflation model by putting it again within the confidence limits of current observational data sets. However, forthcoming CMB experiments like BICEP3 and LiteBIRD which expect to put
stronger constraints on the tensor-to-scalar ratio, may eventually rule out chaotic
quadratic inflation in this particular class of model. For completeness, in the case of other monomial models with $n=4/3$, $n=1$, and $n=2/3$, we found that under a non-minimal coupling to torsion, these models are also in good agreement with observations, with their predictions superimposed over the 
$68\%$ and $95\%$ C.L. regions of the 2018 Planck data.

As it has been extensively studied, there are several mechanism which account in lowering the predictions for the tensor-to-scalar ratio $r$ in a potentially driven single-field scenario. Besides the possibility of a non-minimal coupling to gravity, as we did it previously, one could also include a non-canonical kinetic term to the scalar field action as in $k$-inflation \cite{ArmendarizPicon:1999rj}, and so, getting a subluminal inflaton speed of sound  \cite{Garriga:1999vw}, or either incorporating a damping term due to dissipation, as in warm inflation \cite{Berera:1995ie,Bastero-Gil:2016qru}, or by considering a non-linear field-interaction in the form of the Galileon term $(\partial{\phi})^2 \Box \phi$, which generally works to slow down the evolution of the field and therefore allowing a reduction in the value of $r$ \cite{Ohashi:2012wf}. Thus, in the second part of this paper, we have followed this latter interesting direction. 

Unlike the case with only non-minimal coupling to torsion, whose background equations were
integrated analytically under the slow-roll approximation, in the presence of both non-minimal coupling and Galileon self-interaction, it is not possible to obtain analytical solutions for the background equations, instead we integrate these numerically. In the case of a power-law potential
$V(x)=\lambda x^n/n$ with $n=2$ and the non-minimal coupling function $F(x)=1+\xi x^2/2$, we find
two branches of solutions for $x^{(1)}$ and $x^{(2)}$ (FIG \ref{FIG_Galileon_x}). At first, for solution $x^{(1)}$, which is defined for $\zeta\equiv (\gamma/M_{pl})\times \lambda< 0.13$, with $\gamma$ being the Galileon coupling parameter, we found that the predictions of the model at level of $n_s-r$
plane are within the $95\%$ C.L. contour. Accordingly, for $N=70$, we have found the constraints $7.2\times 10^{-4}\lesssim \xi\lesssim 3.2\times 10^{-3}$, $\lambda/ M_{pl}^4 \sim 10^{-11}$ ($m_{\phi}/M_{pl}\sim 10^{-6}$), and $1\lesssim \gamma/ M_{pl}^{-3} \lesssim 10^{9}$. Therefore, at $N=70$, the tensor-to-scalar ratio takes values in the range $0.028\lesssim r\lesssim 0.069$, which are seen to be reduced as compared to those obtained when it is only considered a non-minimal coupling term. In addition, the required number of $e$-folds is not reduced and the parameters $\xi$ and $\lambda$ are kept in the same order of magnitude as before.

Secondly, when the solution $x^{(2)}$ is studied, which is defined for $\zeta > 0.13$, we found that the predictions in the $n_s-r$ plane are consistent with current observational bounds by Planck 2018 data. This allows us to obtain a lower bound for $\xi$ in the form $\xi\gtrsim 3.0\times 10^{-3}$ for $N=60$, and $\xi\gtrsim 1.4\times 10^{-3}$ for $N=70$. In view of this, it is not possible to obtain
an upper bound for $\xi$ and $\zeta$ by means the $n_s-r$ plane, which could leads us to arbitrarily large values of parameters $\lambda$ and $\gamma$. In order to overcome this problem, one may, for instance, resort to the framework of effective field theories. By considering slow-roll inflation
as a low-energy effective theory, one has that the UV cut-off is given by the Planck scale, that is to say, $\Lambda\lesssim M_{pl}$, while the minimal cut-off is fixed by the inflationary Hubble scale, $\Lambda\gtrsim H$, with the masses of the light fields satisfying the requirement $m\lesssim H$ \cite{Baumann:2014nda}. From our numerical results we found that $H/M_{pl} \sim 10^{-5}$, and for $\Lambda\simeq \gamma^{-1/3}$, it is straight to obtain the condition $\gamma/M_{pl}^{-3}\lesssim 10^{15}$, and so also, $m_{\phi}/ M_{pl}\lesssim 10^{-5}$, or equivalently, $\lambda/M_{pl}^4\lesssim 10^{-10}$. In Table \ref{Table2}, it can be seen that the constraint imposed on $m_{\phi}$ is stronger than those imposed on the Galileon coupling $\gamma$, when determining the upper limit for the non-minimal coupling parameter $\xi$. Therefore, the constraint found on $m_{\phi}$ allows us to set the upper bound on $\xi$ to be $\xi \lesssim 1$, with the better predictions for $r$ in the ranges $0.038\lesssim r\lesssim 0.070$ for $N=60$, and $0.022\lesssim r\lesssim 0.063$ for $N=70$.

Furthermore, an even tighter constraint from above may be imposed to the non-minimal coupling parameter whether we assume that the dominance of Galileon self-interaction over the standard kinetic cannot be extended beyond the end of slow-roll inflation. The reason for this is because the oscillatory regime of inflaton could be break down, and then spoiling the reheating period after inflation \cite{Ohashi:2012wf}. For solution $x^{(2)}$ we found that the transition from the regime $\mathcal{A}> 1$ to the regime $\mathcal{A}< 1$ occurs before the end of inflation providing that $\zeta \lesssim 2$. Hence, the non-minimal coupling parameter is strongly constrained from above to be $\xi\lesssim 0.008$ for $N=60$,  and $\xi\lesssim 0.009$ for $N=70$. Thus, we obtain the bounds $10^{-11} \lesssim \lambda/M_{pl}\lesssim 10^{-10}$ and $10^{9} \lesssim \gamma/M_{pl}^{-3}\lesssim 10^{10}$, being that for $r$ it is found $r\sim 0.042$ for $N=60$ and $r\sim 0.024$ for $N=70$, which is consistent with our previous constraints obtained from the framework of effective field theories. Then, chaotic quadratic inflation in
our scalar-torsion gravity scenario can be reconciled with current Planck data even within the $68\%$ C.L. contour region. In despite of this, next generation CMB experiments
such as BICEP3 or LiteBIRD are expected to put stronger constraints, making possible to either support this model or rule it out in near future.

As a final remark, it is important to note that a more detailed studied of the post-inflationary phase should be performed, in order to obtain a better picture of the dynamics of reheating in these models with non-minimal coupling to torsion. We hope to be able to address this point in a future work.




\appendix

\section{Appendix: Tensor pertubations}\label{Appendix}

A more general case of scalar-torsion gravity is

\begin{equation}
S=\int d^4x e \left[\frac{M_{pl}^2}{2} F(\phi, X) T+P(\phi, X)- G(\phi, X)\Box\phi \right], 
\end{equation}

but this action, like scalar-curvature gravity, is consistent with GW170817 and GRB170817A
only for a particular case. We can probe this through the analysis of the tensor pertubations,
\be
S_{T}=\sum_{\lambda} \int{dt d^3x a^3 Q_{T}\left[\dot{h}_{\lambda}^2-\frac{c_{T}^2}{a^2} \left(\partial h_{\lambda}\right)^2\right]},
\ee where two polarization states are given by $\lambda=+,\times$. The quantity $Q_{T}$ is defined by
\be
Q_{T}=  \dfrac{M^{2}_{pl}}{8} \left( F + 2 X F_X \right),
\ee
and the squared tensor propagation speed is
\be
c_{T}^2= \dfrac{F}{F + 2 X F_X},
\ee where we can see that $c_T^2$ is equivalent to 1 only for the particular case $F=F(\phi)$. Therefore, we must restrict to the action given by \eqref{action1}.


\def\tablename{Table}%
\begin{table*}[!ht]
\centering
\begin{center}
\renewcommand{\tabcolsep}{1.2pc} 
\renewcommand{\arraystretch}{1} 
\begin{tabular}{| c | c | c | c | c |}\hline\cline{1-5}
$\xi$ &   $\zeta$  & $10^{11}\lambda/M_{pl}^4$  & $ 10^{6}m_{\phi}/M_{pl}$   &  $10^{-10}M_{pl}^3\gamma$ 
\\\cline{1-5} 
 $0.0010$ & $(1\times 10^{-10},0.05)$ & $(1.42, 2.35)$ & $(3.76, 4.85)$ &  $(7.06 \times 10^{-10},0.213)$ \\\cline{1-5}
 $0.0015$ & $(0.0125,0.13)$ & $(1.26, 2.73)$ & $(3.56, 5.22)$ &  $(0.0989,0.477)$ \\\cline{1-5} 
 $0.0020$ & $(0.035,0.13)$  & $(1.28, 2.36)$ &  $(3.58, 4.86)$ &   $(0.273,0.551)$\\\cline{1-5}
 $0.0025$ & $(0.069, 0.13)$ & $(1.37, 1.99)$ & $(3.71, 4.46)$ & $(0.502,0.654)$\\\cline{1-5}
 \hline
\end{tabular}
\end{center}
\caption{Summary on the parameters $\zeta$, $\lambda$, and $\gamma$, for some values of the parameter $\xi$, in the case of solution $x^{(1)}$, for $n=2$ and $N=70$.} 
\label{Table1}
\end{table*}

\def\tablename{Table}%
\begin{table*}[!ht]
\centering
\begin{center}
\renewcommand{\tabcolsep}{0.2pc} 
\renewcommand{\arraystretch}{1.5} 
\begin{tabular}{| c | c | c | c | c | c |}\hline\cline{1-6}
 $\xi$ &   $\zeta$  & $10^{11}\lambda/M_{pl}^4  $ & $ 10^{6}m_{\phi}/M_{pl}$ &  $10^{-10}M_{pl}^3\gamma$ & $N$
\\\cline{1-6} 
 $0.002$ & $(0.14,0.45)$  & $(2.44, 4.04)$ & $(4.94, 6.35)$ &  $(0.57, 1.11)$ & 70 \\\cline{1-6} 
 $0.004$ & $(0.36, 1.45)$   & $(3.53, 7.49)$ &  $(5.94, 8.65)$ &   $(1.02, 1.94)$ & 60   \\\cline{2-6}
  &  $(0.24,2.9)$  & $(1.70, 7.59)$ &  $(4.13, 8.71)$ &   $(1.41, 3.82)$ & 70  \\\cline{1-6}
 $0.006$ & $(0.95, 4.2)$ & $(4.61, 10.96)$  & $(6.79, 10.47)$ &  $(2.06, 3.83)$ & 60   \\\cline{2-6}
 & $(0.68,8)$ & $(2.27, 10.99)$  & $(4.77, 10.48)$ &  $(2.99, 7.28)$ & 70   \\\cline{1-6}
 $0.008$ & $(1.95,9)$ & $(5.85, 14.59)$  & $(7.65, 12.08)$ & $(3.33, 6.17)$ & 60  \\\cline{2-6} 
  & $(1.42,16)$ & $(2.89, 14.22)$  & $(5.38, 11.93)$ & $(4.91, 11.25)$ & 70  \\\cline{1-6} 
 $0.01$ & $(3.3,15)$ &  $(6.95, 17.56)$ &  $(8.34, 13.25)$  &  $(4.75, 8.54)$ & 60   \\\cline{2-6}
   & $(2.5,30)$ &  $(3.52, 18.24)$ &  $(5.94, 13.50)$  &  $(7.10, 16.45)$ & 70   \\\cline{1-6}
 $0.03$  & $(51, 250)$ & $(19.36, 52.75)$ & $(13.91, 22.97)$  & $(26.34, 47.39)$   & 60   \\\cline{2-6}
   & $(39, 450)$ & $(9.78, 52.31)$ & $(9.89, 22.87)$  & $(39.86, 86.03)$   & 70   \\\cline{1-6}
 $0.05$ & $(180, 900)$ & $(31.48, 87.54)$ &  $(17.74, 29.59)$ & $(57.17, 102.82)$  & 60  \\\cline{2-6}
  & $(140, 1650)$ & $(16.11, 87.75)$ &  $(12.69, 29.62)$ & $(86.90, 288.04)$  & 70  \\\cline{1-6}
 $0.1$ & $(1000, 5000)$ & $(61.53, 172.56)$ & $(24.80, 41.54)$ & $(1.63 \times 10^{2}, 2.90 \times 10^{2})$ & 60 \\\cline{2-6}
  & $(790, 9500)$ & $(31.84, 176.45)$ & $(17.84, 42.01)$ & $(2.48 \times 10^{2}, 5.38 \times 10^{2})$ & 70 \\\cline{1-6}
$1$ & $(3.2 \times 10^5, 1.6 \times 10^6)$ & $(6.17 \times 10^2, 1.73 \times 10^3)$ & $(78.52, 131.50)$ & $(5.19 \times 10^3, 9.25 \times 10^3)$ & 60 \\\cline{2-6}
 & $(2.5\times 10^5, 2.9\times 10^6)$ & $(3.16 \times 10^2, 1.73 \times 10^3)$ & $(56.22, 131.42)$ & $(7.91 \times 10^{3},1.68 \times 10^{4})$ & 70 \\\cline{1-6}

 $10$ & $(1.02 \times 10^8, 5.3 \times 10^8)$ & $(6.20 \times 10^3, 1.77 \times 10^4)$ & $(249.01, 420.97)$ & $(1.65 \times 10^5,2.99 \times 10^5)$ & 60 \\\cline{2-6}
  & $(7.85 \times 10^7, 9 \times 10^8)$ & $(3.13 \times 10^3, 1.71 \times 10^4)$ & $(177.03, 413.49)$ & $(2.50 \times 10^5,5.26 \times 10^5)$ & 70 \\\cline{1-6}
 \hline
\end{tabular}
\end{center}
\caption{Summary of the bounds on the parameters of the model $\zeta$, $\lambda$, and $\gamma$, for some values of the parameter $\xi$, for the solution $x^{(2)}$, in the case of $n=2$, being that for the $e$-folds number we have considered the values, $N=60$ and $N=70$.} 
\label{Table2}
\end{table*}


\acknowledgments
G. Otalora acknowldeges DI-VRIEA for financial support through Proyecto Postdoctorado $2019$ VRIEA-PUCV. M. Gonzalez-Espinoza acknowledges support from a PUCV doctoral scholarship. N.V. acknowledges support from the Fondecyt de Iniciaci\'on
project N$^o$ 11170162.

\clearpage


\bibliographystyle{spphys}       
\bibliography{bio}   



\end{document}